\documentclass[prd,reprint,floats,amsmath,amssymb,showpacs,aps,superscriptaddress,notitlepage]{revtex4-1}
\usepackage{graphicx}
\usepackage{amssymb}
\usepackage{epstopdf}
\usepackage{color}

\usepackage{dsfont}
\usepackage{hyperref}
\usepackage{graphicx}
\usepackage{graphics}
\usepackage{epsfig}
\usepackage{bm}
\usepackage{longtable}
\usepackage{verbatim}
\usepackage{yfonts}
\usepackage{psfrag}
\usepackage{url}
\usepackage{subfigure}
\usepackage{color}
\input{epsf}
\usepackage{thumbpdf}
\usepackage{pdfsync}
\usepackage{graphicx}
\usepackage{enumerate}

\def\be{\begin{equation}}
\def\en{\end{equation}}

\def\msun{\,{\rm M_\odot}}
\def\0{\hphantom{0}}

\begin{document}

\title{Octahedron configuration for a displacement noise-canceling gravitational wave detector in space}
\author{Yan Wang}
\email{yan.wang@aei.mpg.de}
\author{David Keitel}
\email{david.keitel@aei.mpg.de}
\affiliation{Max-Planck-Institut f\"ur Gravitationsphysik (Albert-Einstein-Institut), Callinstra{\ss}e 38, 30167 Hannover, Germany}
\author{Stanislav Babak}
\email{stba@aei.mpg.de}
\affiliation{Max-Planck-Institut f\"ur Gravitationsphysik (Albert-Einstein-Institut), Am M\"uhlenberg 1, 14476 Golm, Germany}
\author{Antoine Petiteau}
\affiliation{Max-Planck-Institut f\"ur Gravitationsphysik (Albert-Einstein-Institut), Am M\"uhlenberg 1, 14476 Golm, Germany}
\affiliation{APC, Universit\'{e} Paris Diderot,  Observatoire de Paris, Sorbonne Paris Cit\'{e}, 10 rue Alice Domon et L\'{e}onie Duquet, 75205 Paris Cedex 13, France}
\author{Markus Otto}
\email{markus.otto@aei.mpg.de}
\author{Simon Barke}
\author{Fumiko Kawazoe}
\author{Alexander Khalaidovski}
\author{Vitali M\"uller}
\author{Daniel Sch\"utze}
\author{Holger Wittel}
\author{Karsten Danzmann}
\affiliation{Max-Planck-Institut f\"ur Gravitationsphysik (Albert-Einstein-Institut), Callinstra{\ss}e 38, 30167 Hannover, Germany}
\author{Bernard F. Schutz}
\affiliation{Max-Planck-Institut f\"ur Gravitationsphysik (Albert-Einstein-Institut), Am M\"uhlenberg 1, 14476 Golm, Germany}

\date{\today}

\begin{abstract}
We study for the first time a three-dimensional octahedron constellation for a space-based gravitational wave detector,
which we call the Octahedral Gravitational Observatory (OGO).
With six spacecraft the constellation is able to remove laser frequency noise and acceleration disturbances from the
gravitational wave signal without needing LISA-like drag-free control, thereby simplifying the payloads and placing
less stringent demands on the thrusters. We generalize
LISA's time-delay interferometry to displacement-noise free interferometry (DFI) by deriving
a set of generators for those combinations of the data streams  that cancel laser and
acceleration noise.
However, the three-dimensional configuration makes orbit selection complicated.
So far, only a halo orbit near the Lagrangian point L1 has been found to be stable enough, and this allows only short arms up to 1400\,km.
We derive the sensitivity curve of OGO with this arm length, resulting in a peak sensitivity of about 2$\times10^{-23}\,\mathrm{Hz}^{-1/2}$ near 100\,Hz.
We compare this version of OGO to the
present generation of ground-based detectors and to some future detectors. We also investigate the scientific
potentials of such a detector, which include observing gravitational waves from compact binary
coalescences, the stochastic background and pulsars as well as the possibility to test alternative
theories of gravity.
We find a mediocre performance level for this short-arm-length detector, between those of initial and advanced ground-based detectors.
Thus, actually building a space-based detector of this specific configuration does not seem very efficient.
However, when alternative orbits that allow for longer detector arms can be found, a detector with much improved science output could be constructed using the octahedron configuration and DFI solutions demonstrated in this paper.
Also, since the sensitivity of a DFI detector is limited mainly by shot noise, we discuss how the overall sensitivity could be improved
by using advanced technologies that reduce this particular noise source.
\end{abstract}



\maketitle

\section{Introduction}
 The search for gravitational waves (GWs) has been carried out for more than a decade by ground-based detectors.
 Currently, the LIGO and Virgo detectors are being upgraded using advanced technologies~\cite{aLIGO, adVIRGO}.
 The ground-based detectors are sensitive in quite a broad band from about 10\,Hz to a few kHz. In this band possible GW sources include stellar-mass compact coalescing binaries~\cite{Abadie2010b}, asymmetric core collapse of evolved heavy stars~\cite{FryerNew2011}, neutron stars with a nonzero ellipticity~\cite{Owen2009} and, probably, a stochastic GW background from the early Universe or from a network of cosmic strings~\cite{Allen99, Maggiore00}.

 In addition, the launch of a space-based GW observatory is expected in the next decade, such as the classic LISA mission concept~\cite{LISA} (or its recent modification known as evolved LISA (eLISA) / NGO~\cite{eLISA}), and DECIGO~\cite{Ando2010}.
 LISA has become a mission concept for any heliocentric drag-free configuration that uses laser interferometry
 for detecting GWs.
 The most likely first GW observatory in space will be the eLISA mission, which has an arm length of $10^9$\,m and two arms, with one ``mother'' and two ``daughter'' spacecraft exchanging laser light in a V-shaped configuration to sense the variation of the metric due to passing GWs.

 The eLISA mission aims at mHz frequencies, targeting other sources than ground-based detectors, most importantly supermassive black hole binaries.
 In a more ambitious concept, DECIGO is supposed to consist of a set of four smaller triangles (12 spacecraft in total) in a common orbit, leading to a very good sensitivity in the intermediate frequency region between LISA and advanced LIGO (aLIGO).

 Here we want to present a concept for another space-based project with quite a different configuration from what has been considered before.
 The concept was inspired by a three-dimensional interferometer configuration in the form of an octahedron, first suggested in Ref.~\cite{chen2006} for a ground-based detector, based on two Mach-Zehnder interferometers.

\begin{figure*}[hbt]
  \centering
  \includegraphics[width=\textwidth, keepaspectratio=true]{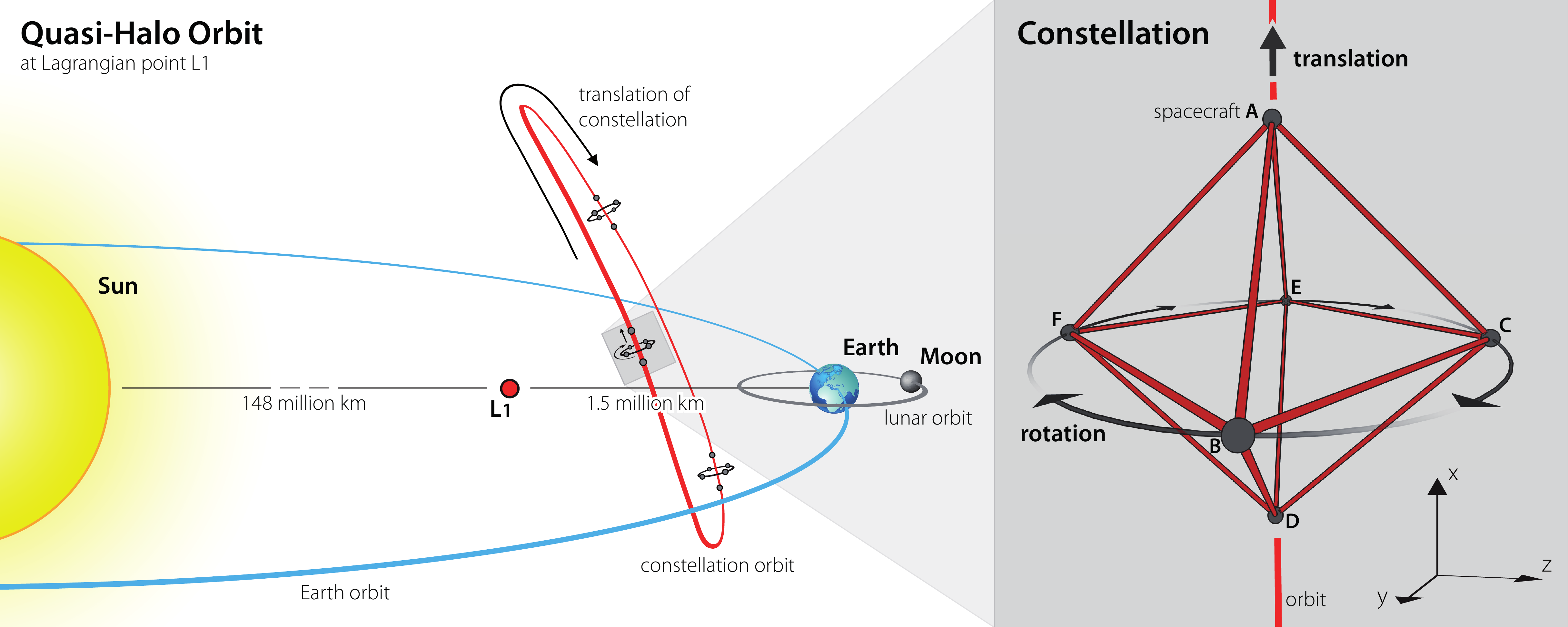}\\
  \caption{Left: Graphical representation of the proposed halo orbit around L1. Right: OGO's spacecraft constellation along the halo orbit, with a radius of 1000\,km and spacecraft separation of $L=\sqrt{2}\,r\approx1400$\,km.}
  \label{F:orbit}
 \end{figure*}

 The main advantage of this setup is the cancellation of timing, laser frequency and displacement noise by combining multiple measurement channels.
 We have transformed this detector into a space-borne observatory by placing one LISA-like spacecraft (but with four telescopes and a single test mass) in each of the six corners of the octahedron, as shown in Fig.~\ref{F:orbit}.
 Therefore, we call this project the \emph{Octahedral Gravitational Observatory} (OGO).

 Before going into the mathematical details of displacement-noise free interferometry (DFI), we first consider possible orbits for a three-dimensional octahedron constellation in Sec.~\ref{S:Orbit}.
 As we will find later on, the best sensitivities of an OGO-like detector are expected at very long arm lengths.
 However, the most realistic orbits we found that can sustain the three-dimensional configuration with stable distances between adjacent spacecraft for a sufficiently long time are so-called ``halo'' and ``quasihalo'' orbits around the Lagrange point L1 in the Sun-Earth system.

 These orbits are rather close to Earth, making a mission potentially cheaper in terms of fuel and communication, and corrections to maintain the formation seem to be reasonably low.
 On the other hand, a constellation radius of only 1000\,km can be supported, corresponding to a spacecraft-to-spacecraft arm length of approximately 1400\,km.

 We will discuss this as the standard configuration proposal for OGO in the following, but ultimately we still aim at using much longer arm lengths. As a candidate, we will also discuss OGO orbits with $2\times10^9\,$m arm lengths in Sec.~\ref{S:Orbit}. However, such orbits might have significantly varying separations and would require further study of the DFI technique in such circumstances.

 The octahedron configuration gives us 24 laser links, each corresponding to a science measurement channel of the distance (photon flight-time) variation between the test masses on adjacent spacecraft.
 The main idea is to use a sophisticated algorithm called \emph{displacement-noise free interferometry} (DFI,~\cite{kawamura2004,chen2006,chenkawa2006}), which proceeds beyond conventional Time-Delay Interferometry techniques (TDI,~\cite{TintoDhurandhar,otto2012}), and in the right circumstances can improve upon them.

 It can cancel both timing noise and acceleration noise when there are more measurements than noise sources.
 In three dimensions, the minimum number of spacecraft for DFI is 6, which we therefore use for OGO: this gives $6-1$ relative timing (clock) noise sources and $3\times 6 = 18$ components of the acceleration noise, so that $24 > 5+18$ and the DFI requirement is fulfilled.
 On the one hand, this required number of links increases the complexity of the detector. On the other hand, it provides some redundancy in the number of shot-noise-only configurations, which could be very useful if one or several links between spacecraft are interrupted.

 After applying DFI, we assume that the dominant remaining noise will be shot noise.
 For the case of an equal-arm-length three-dimensional constellation, we analytically find a set of generators for the measurement channel combinations that cancel simultaneously all timing and acceleration noise.
 We assume that all deviations from the equal-arm configuration are small and can be absorbed into a low-frequency part of the acceleration noise.
 We describe the procedure of building DFI combinations in Sec.~\ref{S:TDI}.
 This will also allow us to quantify the redundancy inherent in the six-spacecraft configuration.
 The technical details of the derivation can be found in Appendix~\ref{S:Appendix}.

 In Sec.~\ref{S:Sens}, we compute the response functions of the octahedron DFI configuration and derive the sensitivity curve of the detector.
 We assume the conservative 1400\,km arm length, a laser power of 10\,W and a telescope diameter of 1\,m, while identical strain sensitivity is achievable for smaller telescopes and higher power.

 Unfortunately, those combinations that cancel acceleration and timing noise also suppress the GW signal at low frequencies.
 This effect shows up as a rather steep slope $\sim f^{2}$ in the response function.

 We present sensitivity curves for single DFI combinations and find that there are in principle 12 such noise-uncorrelated combinations (corresponding to the number of independent links) with similar sensitivity, leading to an improved network sensitivity of the full OGO detector.
 We find that the best sensitivity is achieved around 78\,Hz, in a range similar to that of ground-based detectors.
 The network sensitivity of OGO is better than that of initial LIGO at this frequency, but becomes better than that of aLIGO only below
 10\,Hz. The details of these calculations are presented in Sec.~\ref{sec:transfer_function}.

 At this point, in Sec.~\ref{sec:performance}, we briefly revisit the alternative orbits with a longer arm length, which would result in a sensitivity closer to the frequency band of interest for LISA and DECIGO.
 For this variant of OGO, we assume LISA-like noise contributions (but without spacecraft jitter) and compare the sensitivity of an octahedron detector using DFI with one using TDI, thus directly comparing the effects of these measurement techniques.

 Actually, we find that the $2\times10^9\,$m arm length is close to the point of equal sensitivity of DFI and TDI detectors in the limit of vanishing jitter.
 This implies that DFI would be preferred for even longer arm lengths, but might already become competitive at moderate arm lengths if part of the jitter couples into the displacement noise in such a way that it can also be canceled.

 A major advantage of the OGO concept lies in its rather moderate requirement on acceleration noise, as detailed in Sec.~\ref{sec:feasibility}.
 For other detectors, this limits the overall performance, but in this concept it gets canceled out by the DFI combinations.
 Assuming some improvements in subdominant noise sources, our final sensitivity thus depends only on the shot-noise level in each link.

 Hence, we can improve the detector performance over all frequencies by reducing solely the shot noise.
 This could be achieved, for example, by increasing the power of each laser, by introducing cavities (similar to DECIGO), or with nonclassical (squeezed) states of light.
 We briefly discuss these possibilities in Sec.~\ref{sec:shot_noise_reduction}.

 In Sec.~\ref{S:Sources}, we discuss the scientific potentials OGO would have even using the conservative short-arm-length orbits.
 First, as a main target, the detection rates for inspiraling binaries are higher than for initial LIGO, but fall short of aLIGO expectations.
 However, joint detections with OGO and aLIGO could yield some events with greatly improved angular resolution.
 Second, due to the large number of measurement channels, OGO is good for probing the stochastic background.
 Furthermore, the three-dimensional configuration allows us to test alternative theories of gravity by searching for additional GW polarization modes.
 In addition, we briefly consider other source types such as pulsars, intermediate mass
 ($10^2  < M/M_{\odot} < 10^4$) black hole (IMBH) binaries and supernovae.

 Finally, in Sec.~\ref{S:Summary}, we summarize the description and abilities of the Octahedral Gravitational Observatory  and mention additional hypothetical improvements.
 In this article, we use geometric units, $c = G = 1$, unless stated otherwise.

\section{Orbits}
 \label{S:Orbit}

 The realization of an octahedral constellation of spacecraft depends on the existence of suitable orbits.
 Driving factors, apart from separation stability, are assumed to be (i) fuel costs in terms of velocity $\Delta v$ necessary to deploy and maintain the constellation of six spacecraft, and (ii) a short constellation-to-Earth distance, required for a communication link with sufficient bandwidth to send data back to Earth.
 As described in the introduction, OGO features a three-dimensional satellite constellation.
 Therefore, using heliocentric orbits with a semimajor axis $a = 1$\,AU similar to LISA would cause a significant drift of radially separated spacecraft and is in our opinion not feasible.

 However, in the last decades orbits in the nonlinear regime of Sun/Earth-Moon libration points L1 and L2 have been exploited, which can be reached relatively cheaply in terms of fuel~\cite{Gomez1993}.
 A circular constellation can be deployed on a torus around a halo L1 orbit.
 The radius is limited by the amount of thrust needed for keeping the orbit stable.
 A realistic $\Delta v$ for orbit maintenance allows a nominal constellation radius of $r=1000$\,km~\cite{Howell1999}.
 We assume the spacecraft B, C, E and F in Fig.~\ref{F:orbit} to be placed on such a torus, whereby the out-of-plane spacecraft A and D will head and trail on the inner halo.
 The octahedron formation then has a base length $L=\sqrt{2}\,r\approx1400$\,km.
 The halo and quasihalo orbits have an orbital period of roughly 180 days and the whole constellation rotates around the A-D line.

 We already note at this point that a longer baseline would significantly improve the detector strain sensitivity.
 Therefore, we also propose an alternative configuration with an approximate average side length of $2\times10^9\,$m, where spacecraft A and D are placed on a small halo or Lissajous orbit around L1 and L2, respectively. The remaining spacecraft are arranged evenly on a (very) large halo orbit around either L1 or L2.
 However, simulations using natural reference trajectories showed that this formation is slightly asymmetric and that the variations in the arm lengths (and therefore in the angles between the links) are quite large.
 Nevertheless, we will revisit this alternative in Sec.~\ref{sec:performance} and do a rough estimation of its sensitivity. To warrant a full scientific study of such a long-arm-length detector would first require a more detailed study of these orbits.

 Hence, we assume the 1400 km constellation to be a more realistic baseline, especially since the similarity of the spacecraft orbits is advantageous for the formation deployment, because large (and expensive) propulsion modules for each satellite are not required as proposed in the LISA/NGO mission~\cite{NGOYellowBook,LisaYellow}. The $2\times10^9\,$m formation will be stressed only to show the improvement of the detector sensitivity with longer arms.

 Formation flight in the vicinity of Lagrange points L1 and L2 is still an ongoing research topic~\cite{Folta2004}.
 Detailed (numerical) simulations have to be performed to validate these orbit options and to figure out appropriate orbit and formation control strategies. In particular the suppression of constellation deformations using non-natural orbits with correction maneuvers and required $\Delta v$ and fuel consumption needs to be investigated. Remaining deformations and resizing of the constellation will likely require a beam or telescope steering mechanism on the spacecraft.

 In addition, the formation will have a varying Sun-incidence angle, leading to further issues for power supply, thermal shielding and blinding of interferometer arms.
 These points need to be targeted at a later stage of the OGO concept development as well as the effect of unequal arms on the DFI scheme.

\section{Measurements and noise-canceling combinations}
 \label{S:TDI}

 In this section we will show how to combine the available measurement channels of the OGO detector to cancel laser and acceleration noise.

 Each spacecraft of OGO is located at a corner of the octahedron, as shown in Fig.~\ref{F:orbit}, and it exchanges laser light with four adjacent spacecraft.
 We consider interference between the beam emitted by spacecraft $I$ and received by spacecraft $J$ with the local beam in $J$, where $I,J = \{\mathrm{A,B,C,D,E,F}\}$ refer to the labels in Fig.~\ref{F:orbit}.  For the sake of simplicity, we assume a rigid and nonrotating constellation.
 In other words, all arm lengths in terms of light travel time are equal, constant in time and independent of the direction in which the light is exchanged between two spacecraft.
 This is analogous to the first generation TDI assumptions~\cite{TintoDhurandhar}. If the expected  deviations from the equal arm configuration are small, then they can be absorbed into the low-frequency part of the acceleration noise.
 This imposes some restrictions on the orbits and on the orbit correction maneuvers. We also want to note that the
 overall breathing of the constellation (scaling of the arm length) is not important if the breathing time scale
 is significantly larger than the time required for the DFI formation, which is usually true. All calculations below
 are valid if we take the arm length at the instance of DFI formation, which is the value that affects the sensitivity of the detector.

 The measurement of the fractional frequency change for each link is then given by
 \begin{equation}
  s^{\mathrm{tot}}_{IJ} = h_{IJ} + b_{IJ} + \mathcal{D} p_{I} - p_{J} + \mathcal{D} \left( \vec{a}_{I}\cdot\hat{n}_{IJ} \right) - \left( \vec{a}_{J}\cdot\hat{n}_{IJ} \right) \, ,
  \label{E:Mes1Gen}
 \end{equation}
 where we have neglected the factors to convert displacement noise to optical frequency shifts. Here, we have the following:

 \begin{enumerate}[(i)]
  \item $h_{IJ}$ is the influence of gravitational waves on the link $I \rightarrow J$,
  \item $b_{IJ}$ is the shot noise (and other similar noise sources at the photo detector and phase meter of spacecraft $J$) along the link $I \rightarrow J$.
  \item $p_{I}$ is the laser noise of spacecraft $I$.
  \item $\vec{a}_{I}$ is the acceleration noise of spacecraft $I$.
  \item $\hat{n}_{IJ} = ( \vec{x}_{J} - \vec{x}_{I}) / L$ is the unit vector along the arm $I \rightarrow J$ (with length $L$).
        Hence, the scalar product $\vec{a}_I \cdot \hat{n}_{IJ}$ is the acceleration noise of spacecraft $I$ projected onto the arm characterized by the unit vector $\hat{n}_{IJ}$.
 \end{enumerate}
 This is similar to TDI considerations, but in addition to canceling the laser noise $p_I$, we also want to eliminate the influence of the acceleration noise, that is all terms containing $a_I$.
 Following Ref.~\cite{TintoDhurandhar}, we have introduced a delay operator  $\mathcal{D}$, which acts as
 \begin{equation}
  \mathcal{D} y(t) = y( t - L) \, .
  \label{E:Delay}
 \end{equation}
 Note that we use a coordinate frame associated with the center of the octahedron, as depicted in Fig.~\ref{F:orbit}.

 The basic idea is to find combinations of the individual measurements (Eq.~\ref{E:Mes1Gen}) which are free of acceleration noise $\vec{a}_{I}$ and laser noise $p_{I}$.
 In other words, we want to find solutions to the following equation:
 \begin{equation}
  \sum_{\textrm{all} \ IJ \ \textrm{links}} q_{IJ} \ s_{IJ} = 0\,.
  \label{E:GenCondNoiseNull}
 \end{equation}
 In Eq.~(\ref{E:GenCondNoiseNull}), $q_{IJ}$ denotes an unknown function of delays $\mathcal{D}$ and $ s_{IJ}$ contains only the noise we want to cancel:
 \begin{eqnarray}
  s_{IJ} &\equiv& s^{\mathrm{tot}}_{IJ}(b_{IJ} = h_{IJ} = 0) \nonumber \\
  &=& \mathcal{D} p_{I} - p_{J} + \mathcal{D} \left( \vec{a}_{I}\cdot\hat{n}_{IJ} \right) - \left( \vec{a}_{J}\cdot\hat{n}_{IJ} \right).
  \label{E:MesN}
 \end{eqnarray}
 If a given $q_{IJ}$ is a solution, then $f(\mathcal{D})q_{IJ}$ is also a solution, where $f(\mathcal{D})$ is a polynomial function (of arbitrary order) of delays.  The general method for finding generators of the solutions for this equation is described in Ref.~\cite{TintoDhurandhar} and we will follow it closely.

 Before we proceed to a general solution for Eq.~(\ref{E:GenCondNoiseNull}), we can check that the solution corresponding to Mach-Zehnder interferometers suggested in Ref.~\cite{chen2006} also satisfies
 Eq.~(\ref{E:GenCondNoiseNull}):
 \begin{subequations}
 \begin{align}
  Y_1 &= [ \,( s_{CD} + \mathcal{D} s_{AC} ) - ( s_{CA} + \mathcal{D} s_{DC} ) + ( s_{FD} + \mathcal{D} s_{AF} ) \nonumber \\
  &- ( s_{FA} + \mathcal{D} s_{DF} ) \, ] - [ \, ( s_{BD} + \mathcal{D} s_{AB} ) - ( s_{BA} + \mathcal{D} s_{DB} ) \nonumber \\
  & + ( s_{ED} + \mathcal{D} s_{AE} ) - ( s_{EA} + \mathcal{D} s_{DE} )\, ]  \,.
 \end{align}
 Using the symmetries of an octahedron, we can write down two other solutions:
 \begin{align}
  Y_2 &=  [\, (s_{CE} + \mathcal{D} s_{BC}) - (s_{CB} + \mathcal{D} s_{EC}) + (s_{FE} + \mathcal{D} s_{BF}) \nonumber \\
  &- (s_{FB} + \mathcal{D} s_{EF})\,] - [\,(s_{AE} + \mathcal{D} s_{BA}) - (s_{AB} + \mathcal{D} s_{EA})\nonumber\\
  & + (s_{DE} + \mathcal{D} s_{BD}) - (s_{DB} + \mathcal{D} s_{ED}) \,]\, , \\
        & \nonumber \\
  Y_3 &=  [\,(s_{DF} + \mathcal{D} s_{CD}) - (s_{DC} + \mathcal{D} s_{FD}) + (s_{AF} + \mathcal{D} s_{CA}) \nonumber\\
   &- (s_{AC} + \mathcal{D} s_{FA})\,] - [\,((s_{EF} + \mathcal{D} s_{CE}) - (s_{EC} + \mathcal{D} s_{FE})\nonumber\\
   &+ (s_{BF} + \mathcal{D} s_{CB}) - (s_{BC} + \mathcal{D} s_{FB}) \,]\,.
 \end{align}
\end{subequations}
 We can represent these solutions as 24-tuples of coefficients for the delay functions $q_{IJ}$:
 \begin{subequations}
 \begin{eqnarray}
  q_1 &= & \{ 1, 1, -1, -1, -1, -1, 1, 1, -\mathcal{D}, \mathcal{D}, 0, 0, -\mathcal{D},
\mathcal{D}, 0, 0, \nonumber\\
& & \mathcal{D},-\mathcal{D}, 0, 0, \mathcal{D}, -\mathcal{D}, 0, 0 \} \, , \\
  q_2 &= &  \{ -\mathcal{D}  , \mathcal{D}   , 0   , 0   , -\mathcal{D}  , \mathcal{D}   , 0   , 0   , 1   , 1   , -1  , -1  , -1  , -1  , 1   , 1,  \nonumber\\
  & & 0   , 0   , \mathcal{D}   , -\mathcal{D}  , 0   , 0   , \mathcal{D}   , -\mathcal{D} \} \, , \\
  q_3 &= &  \{ 0   , 0   , \mathcal{D}   , -\mathcal{D}  , 0   , 0   , \mathcal{D}   , -\mathcal{D}  , 0   , 0   , -\mathcal{D}  , \mathcal{D}   , 0   , 0   , -\mathcal{D}  , \mathcal{D}, \nonumber\\
  & &    -1  , -1  , 1   , 1   , 1   , 1   , -1  , -1 \} \, .
 \end{eqnarray}
 \end{subequations}
 The order used in the 24-tuples is $\{ BA$, $EA$, $CA$, $FA$, $BD$, $ED$, $CD$, $FD$, $AB$, $DB$, $CB$, $FB$, $AE$, $DE$, $CE$, $FE$, $AC$, $DC$, $BC$, $EC$, $AF$, $DF$, $BF$, $EF \}$, so that,
 for example, the first entry in $q_1$  represents the $s_{BA}$ coefficient in the $Y_1$ equation.

 These particular solutions illustrate that not all links are used in producing a
 DFI stream. Multiple zeros in the equations for $q_1, q_2, q_3$ above indicate those links which do not
 contribute to the final result, and each time we use only 16 links. We will come back to the issue of ``lost links''
 when we discuss the network sensitivity.

 In the following, we will find generators of all solutions.
 The first step is to use Gaussian elimination (without division by delay operators) in Eq.~(\ref{E:GenCondNoiseNull}), and as a result, we end up with a single (master) equation which we need to solve:
 \begin{eqnarray}
  0 &=& (\mathcal{D}-1)^2 q_{BC} + (\mathcal{D}-1)\mathcal{D} q_{CE} + (1-\mathcal{D})(\mathcal{D}-1)\mathcal{D} q_{DB} \nonumber \\
  &+& (\mathcal{D}-1)((1-\mathcal{D})\mathcal{D}-1) q_{DC} \nonumber\\
  &+& (\mathcal{D}-1) q_{DF} + (\mathcal{D}-1) q_{EF} \, . \label{E:TDIAcc_FinalEq}
 \end{eqnarray}
 In the next step, we want to find the so-called ``reduced generators'' of Eq.~(\ref{E:TDIAcc_FinalEq}), which correspond to the reduced set
$( q_{BC}, q_{CE}, q_{DB}, q_{DC}, q_{DF}, q_{EF} )$.
 For this we  need to compute the Gr\"obner basis~\cite{Buchberger1970}, a set generating the polynomial ideals $q_{IJ}$.
 Roughly speaking, the Gr\"obner basis is comparable to the greatest common divisor of $q_{IJ}$.
 Following the procedure from Ref.~\cite{TintoDhurandhar}, we obtain seven generators:
 \begin{widetext}
  \begin{subequations}
 \begin{eqnarray}
  \label{E:S1}
  S_1 &=& \{ 0, \mathcal{D}^2 + \mathcal{D}, 0,  - \mathcal{D} - \mathcal{D}^2, 1 - \mathcal{D},\mathcal{D}^2 + 1,    -1 + \mathcal{D}, -1 - \mathcal{D}^2, \mathcal{D} - \mathcal{D}^2, 0, -\mathcal{D},\mathcal{D}^2, -\mathcal{D}^2 - 1, -\mathcal{D} - 1, 1, \nonumber \\
          & &  1 + \mathcal{D} + \mathcal{D}^2, -\mathcal{D} + \mathcal{D}^2,0, \mathcal{D}, -\mathcal{D}^2, \mathcal{D}^2 + 1, 1 + \mathcal{D}, -1, -\mathcal{D} - \mathcal{D}^2 - 1 \},\\
  & &\nonumber \\
  S_2 &=& \{ \mathcal{D} + 1, \mathcal{D} + 1, -\mathcal{D} -1,-\mathcal{D}-1, -1+\mathcal{D},\mathcal{D}-1, 1-\mathcal{D}, 1 - \mathcal{D}, -2\mathcal{D},0,\mathcal{D},\mathcal{D},-2\mathcal{D},0, \mathcal{D},\mathcal{D}, 2\mathcal{D}, 0, -\mathcal{D}, \nonumber \\
         & & -\mathcal{D},2\mathcal{D},0,-\mathcal{D},-\mathcal{D}\} , \\
  & &\nonumber \\
  S_3 &=& \{ 0, \mathcal{D}, -\mathcal{D}, 0, - 1,  \mathcal{D} - 1, 1- \mathcal{D},1, 1 - \mathcal{D}, 1, -1 + \mathcal{D}, -1, -\mathcal{D},0, \mathcal{D}, 0,  \mathcal{D}, 0, 0, -\mathcal{D}, \mathcal{D} -1,-1, 1, -\mathcal{D} + 1 \},\\
  & &\nonumber \\
  S_4 &=& \{ \mathcal{D}, -\mathcal{D} + \mathcal{D}^2, \mathcal{D}, -\mathcal{D}-\mathcal{D}^2, 2, -2\mathcal{D}+\mathcal{D}^2+2, -2+2\mathcal{D}, -2-\mathcal{D}^2,2\mathcal{D} - 2 -\mathcal{D}^2,-2,2 - 2\mathcal{D},2+\mathcal{D}^2,\mathcal{D} - \mathcal{D}^2, \nonumber \\
          & & -\mathcal{D},-\mathcal{D},\mathcal{D} + \mathcal{D}^2,-2\mathcal{D} + \mathcal{D}^2,0,0,2\mathcal{D} - \mathcal{D}^2,-\mathcal{D} + \mathcal{D}^2 + 2, 2 + \mathcal{D},-2-\mathcal{D},\mathcal{D} - \mathcal{D}^2 - 2\} , \\
  & &\nonumber \\
  S_5 &=& \{  0, \mathcal{D}^2 + \mathcal{D},  -\mathcal{D}^2, - \mathcal{D}, 1 - \mathcal{D}, \mathcal{D}^2  + 1, \mathcal{D} - \mathcal{D}^2  - 1, -1, \mathcal{D} - \mathcal{D}^2, 0,  -\mathcal{D} + \mathcal{D}^2, 0, -1 - \mathcal{D}^2, -\mathcal{D} - 1, 1 + \mathcal{D}^2, \nonumber \\
          & & 1 + \mathcal{D}, \mathcal{D}^2, \mathcal{D}, 0, -\mathcal{D}^2 - \mathcal{D}, -\mathcal{D} + \mathcal{D}^2 +1, 1, \mathcal{D} - 1, -1 - \mathcal{D}^2\},\\
  & & \nonumber \\
  S_6 &=& \{ \mathcal{D} + 2 + \mathcal{D}^2,\mathcal{D}+\mathcal{D}^3+2,-\mathcal{D} + \mathcal{D}^2 - 2, -\mathcal{D} - 2 - 2\mathcal{D}^2 - \mathcal{D}^3,-2 + 2\mathcal{D},2\mathcal{D} - \mathcal{D}^2 + \mathcal{D}^3 - 2,\nonumber \\
          & & -2\mathcal{D} + 2\mathcal{D}^2 + 2,2 - 2\mathcal{D} - \mathcal{D}^2 - \mathcal{D}^3,\mathcal{D}^2 - 4\mathcal{D} - \mathcal{D}^3, 0, 2\mathcal{D} -2\mathcal{D}^2,2\mathcal{D} + \mathcal{D}^2 + \mathcal{D}^3,-3\mathcal{D} - \mathcal{D}^3,\mathcal{D} - \mathcal{D}^2, \nonumber \\
          & &\mathcal{D} - \mathcal{D}^2, 2\mathcal{D}^2 + \mathcal{D} + \mathcal{D}^3, -\mathcal{D}^2 + 2\mathcal{D} + \mathcal{D}^3,  -2\mathcal{D},0,\mathcal{D}^2 - \mathcal{D}^3,5\mathcal{D} + \mathcal{D}^3,\mathcal{D} + \mathcal{D}^2,-3\mathcal{D} - \mathcal{D}^2,-3\mathcal{D} - \mathcal{D}^3 \} , \\
  & &\nonumber \\
  S_7 &=& \{ 1, 1 + \mathcal{D}, -1, -1 - \mathcal{D}, 0, \mathcal{D}, 0, -\mathcal{D}, -\mathcal{D}, 0, 0, \mathcal{D}, -1 - \mathcal{D}, -1, 1, 1 + \mathcal{D}, \mathcal{D}, 0, 0, -\mathcal{D}, 1 + \mathcal{D}, 1, -1, -1 - \mathcal{D} \}.
  \label{E:S7}
 \end{eqnarray}
   \end{subequations}
 \end{widetext}
 As before, these operators have to be applied to $s_{IJ}$, using the same ordering as given above.
 All other solutions can be constructed from these generators. A detailed derivation of expressions (\ref{E:S1})--(\ref{E:S7}) is given in Appendix~\ref{S:Appendix}.

 Before we proceed, let us make several remarks. The generators found here are not unique, just like in the case of TDI~\cite{TintoDhurandhar}.
 The set of generators does not necessarily form a minimal set, and
 we can only guarantee that the found set of generators  gives us a module of syzygies and can be used to generate other solutions.
 The combinations $S_1$ to $S_7$ applied on 24 raw measurements $s_{IJ}^{\mathrm{tot}}$ eliminate both laser and displacement noise while mostly preserving the gravitational wave signal.
 Note that again in those expressions we do not use all links -- for example, if the link $BA$ is lost due to some reasons, we still
 can use $S_1, S_3, S_5$ to produce DFI streams.

\section{Response functions and sensitivity}
 \label{S:Sens}
 In the previous section we have found generators that produce data streams free of acceleration and laser noise.
 Now we need to apply these combinations to the shot noise and to the GW signal to compute the corresponding response functions.

 \subsection{Shot noise level and noise transfer function}
 We will assume that the shot noise is independent (uncorrelated) in each link and equal in power spectral density, based on identical laser sources and telescopes on each spacecraft.
 We denote the power spectral density of the shot noise by $\widetilde{S}_{\rm sn}$.
 A lengthy but straightforward computation shows that the spectral noise $\tilde{S}_{\mathrm{n},i}$ corresponding to the seven combinations $S_i$, $i=1,\ldots,7$ from Eqs.~(\ref{E:S1}--\ref{E:S7}) is given by
 \begin{subequations}
 \begin{eqnarray}
  \widetilde{S}_{\rm n, 1} &=& \0 16\, \widetilde{S}_{\rm sn} \,\epsilon^2\,( 9 + 2\cos2\epsilon + 3\cos4\epsilon)\,,\\
  \widetilde{S}_{\rm n, 2} &=& 160\, \widetilde{S}_{\rm sn} \,\epsilon^2\,,\\
  \widetilde{S}_{\rm n, 3} &=& \0 48\, \widetilde{S}_{\rm sn} \,\epsilon^2 \,( 2 - \cos2\epsilon)\,,\\
  \widetilde{S}_{\rm n, 4} &=& \0 16\, \widetilde{S}_{\rm sn} \,\epsilon^2 \,(24 -13\cos2\epsilon + 6\cos4\epsilon)\,,\\
  \widetilde{S}_{\rm n, 5} &=& \0 16\, \widetilde{S}_{\rm sn} \,\epsilon^2(\, 9 - 2\cos 2\epsilon + 3\cos 4\epsilon)\,,\\
  \widetilde{S}_{\rm n, 6} &=& \0 16\, \widetilde{S}_{\rm sn} \,\epsilon^2 \,(45 -6\cos2\epsilon+17\cos4\epsilon)\,,\\
  \widetilde{S}_{\rm n, 7} &=& \048\, \widetilde{S}_{\rm sn} \,\epsilon^2 \,(2 + \cos2\epsilon)\,,
 \end{eqnarray}
 \end{subequations}
 where $\epsilon \equiv \omega L/2$, with the GW frequency $\omega$.  In the low frequency limit ($\epsilon \ll 1$), the  noise $\widetilde{S}_{{\rm n }, i}$ for each combination $S_i$ is proportional to $\epsilon^2$.

 Let us now compute the shot noise in a single link.
 We consider for OGO a configuration with LISA-like receiver-transponder links and the following parameters: spacecraft separation $L=1414\,$km, laser wavelength $\lambda=532$\,nm, laser power $P=10\,$W and telescope diameter $D=1\,$m.
 For this arm length and telescope size, almost all of the laser power from the remote spacecraft is received by the local spacecraft.
 Hence, the shot-noise calculation for OGO is different from the LISA case, where an overwhelming fraction of the laser beam misses the telescope~\cite{LisaYellow}.

 For a Michelson interferometer, the sensitivity to shot noise is usually expressed as~\cite{Maggiore_book}
 \begin{equation}
  \sqrt{\widetilde{S}_h(f)}=\frac{1}{2L}\sqrt{\frac{\hbar c \lambda}{\pi P}}\,\,\, [1/\sqrt{\rm Hz}] \, ,
 \end{equation}
 where we have temporarily restored the speed of light $c$ and the reduced Planck constant $\hbar$. Notice that the effect of the GW transfer function is not included here yet.
 For a single link $I \rightarrow J$ of OGO as opposed to a full two-arm Michelson with dual links, $\sqrt{\widetilde{S}_{h,IJ}}$ is a factor of 4 larger.
 However, our design allows the following two improvements:
 (i) Since there is a local laser in $J$ with power similar to the received laser power from $I$, the power at the beam splitter is actually $2P$, giving an improvement of $1/\sqrt{2}$.
     This is also different from LISA, where due to the longer arm length the received power is much smaller than the local laser power.
 (ii) If we assume that the arm length is stable enough to operate at the dark fringe, then we gain another factor of $1/\sqrt{2}$.

 So, we arrive at the following shot-noise-only sensitivity for a single link:
 \begin{equation}
  \label{E:shotnoise1}
  \sqrt{\widetilde{S}_{h ,IJ}(f)}=\frac{1}{L}\sqrt{\frac{\hbar c \lambda}{\pi P}}\,\,\, [1/\sqrt{\rm Hz}] \, .
 \end{equation}

 \subsection{GW signal transfer function and sensitivity}
 \label{sec:transfer_function}
 Next, we will compute the detector response to a gravitational wave signal.
 We assume a GW source located in the direction $\hat{n} = -\hat{k} = \left( \sin\theta\cos\phi, \sin\theta\sin\phi, \cos\theta \right)$ as seen from the detector frame.
 We choose unit vectors
 \begin{equation}
  \hat{u} = \left[
  \begin {array}{c}
    \cos \theta  \cos \phi \\
    \cos  \theta \sin \phi \\
    -\sin \theta
 \end {array} \right],\;\;\;\;\;
  \hat{v} =  \left[
  \begin {array}{c}
   \sin \phi \\
  -\cos \phi \\
  0
  \end {array} \right]
 \end{equation}
orthogonal to $\hat{k}$ pointing tangentially along the $\theta$ and $\phi$ coordinate lines to form a polarization basis. This basis can be described by polarization tensors $\mathbf{e}_+$ and $\mathbf{e}_\times$, given by
\begin{equation}
\mathbf{e}_+ \equiv \hat{u}\otimes \hat{u} - \hat{v}\otimes \hat{v}\,, \ \ \
\mathbf{e}_\times \equiv \hat{u}\otimes \hat{v} + \hat{v}\otimes \hat{u}\,.
\label{E:polbas}
\end{equation}
The single arm fractional frequency response to a GW is~\cite{EW1975}
 \begin{eqnarray}
  h_{IJ} = \frac{H_{IJ} (t - \hat{k}\cdot\vec{x}_{I} - L) - H_{IJ} (t - \hat{k}\cdot\vec{x}_{J} )}{2 \left( 1 - \hat{k}\cdot\hat{n}_{IJ} \right)}  ,
 \label{E:E:GWlink}
 \end{eqnarray}
 where $\vec{x}_I$ is the position vector of the $I$-th spacecraft,
 $L$ the (constant) distance between two spacecraft and
\begin{equation}
   H_{IJ } (t) \equiv h_{+} (t) \ \xi_{+} (\hat{u}, \hat{v}, \hat{n}_{IJ}) + h_{\times} (t) \  \xi_{\times} (\hat{u}, \hat{v}, \hat{n}_{IJ}) \, .
  \end{equation}
 Here $h_{+,\times}(t)$ are two GW polarizations in the basis (\ref{E:polbas}) and

\begin{align}
  \xi_{+}(\hat{u}, \hat{v}, \hat{n}_{IJ})  &\equiv \hat{n}_{IJ}^\textsf{T}\mathbf{e}_+ \hat{n}_{IJ} =
  {\left(\hat{u}\cdot \hat{n}_{IJ} \right)}^{2} - {\left(\hat{v}\cdot \hat{n}_{IJ} \right)}^{2} \, , \nonumber\\
  \xi_{\times}(\hat{u}, \hat{v}, \hat{n}_{IJ}) &\equiv \hat{n}_{IJ}^\textsf{T}\mathbf{e}_\times
  \hat{n}_{IJ} = 2 \left(\hat{u}\cdot \hat{n}_{IJ} \right)\left(\hat{v}\cdot \hat{n}_{IJ} \right) \, .
  \label{E:E:GWStrain}
\end{align}

 In order to find the arm response for arbitrary incident GWs, we can compute the single arm response to a monochromatic GW with Eq.~(\ref{E:E:GWlink}) and then deduce the following general response in the frequency domain,
 \begin{eqnarray}
  h_{IJ}(f) & =& \epsilon \, {\rm sinc}\left[\epsilon(1-\hat{k}\cdot\hat{n}_{IJ})\right]
  \mathrm{e}^{-\mathrm{i}\epsilon[\hat{k}\cdot(\vec{x}_{I} + \vec{x}_{J})/L +1]} \nonumber \\
  & & \ \ \ \times \left[\xi_{+}(\hat{n}_{IJ}) h_{+}(f) + \xi_{\times}(\hat{n}_{IJ}) h_{\times}(f)\right] \,,
 \end{eqnarray}
 where we used the normalized sinc function, conventionally used in signal processing: ${\rm sinc}(x) := {\sin(\pi x)}/(\pi x)$.

 Hence, the transfer function for a GW signal is
 \begin{eqnarray}
  \mathcal{T}_{IJ+,\times}^{\rm GW}(f) &=& \epsilon \, {\rm sinc}\left[\epsilon(1-\hat{k}\cdot\hat{n}_{IJ})\right] \nonumber\\
& & \ \ \ \times\, \mathrm{e}^{-\mathrm{i}\epsilon[\hat{k}\cdot(\vec{x}_{I} + \vec{x}_{J})/L +1]}
  \xi_{+, \times}(\hat{n}_{IJ}) \, .
  \label{Eq:TF}
 \end{eqnarray}
 For the sake of simplicity, we will from now on assume that the GW has ``+'' polarization only.
 This simplification will not affect our qualitative end result.
 Substituting the transfer function for a single arm response into the above 7 generators [Eqs.~(\ref{E:S1})-(\ref{E:S7})], we can get the transfer function $\mathcal{T}_{i}^{\rm GW}$ for each combination. The final expressions are very lengthy and not needed here explicitly.

 Having obtained the transfer function, we can compute the sensitivity for each combination $i=1,\ldots,7$ as
 \begin{equation}
  \label{E:sensitivity1}
  \sqrt{\widetilde{S}_{h,i}} = \sqrt{\frac{\widetilde{S}_{{\rm n},  i}}{\langle(\mathcal{T}^{\rm GW}_i)^2\rangle}} \ ,
 \end{equation}
 where the triangular brackets imply averaging over polarization and source sky location.

 We expect up to 12 independent round trip measurements, corresponding to the number of back-and-forth links between spacecraft.
 It is out of the scope of this work to explicitly find all noise-uncorrelated combinations (similar to the optimal channels $A, E, T$ in the case of LISA~\cite{TintoDhurandhar}).
 However, if we assume approximately equal sensitivity for each combination (which is almost the case for the combinations $S_1,\ldots,S_7$), we expect an improvement in the sensitivity of the whole network by a factor $1/\sqrt{12}$.

 \begin{figure}
  \includegraphics[width=\columnwidth, keepaspectratio=true]{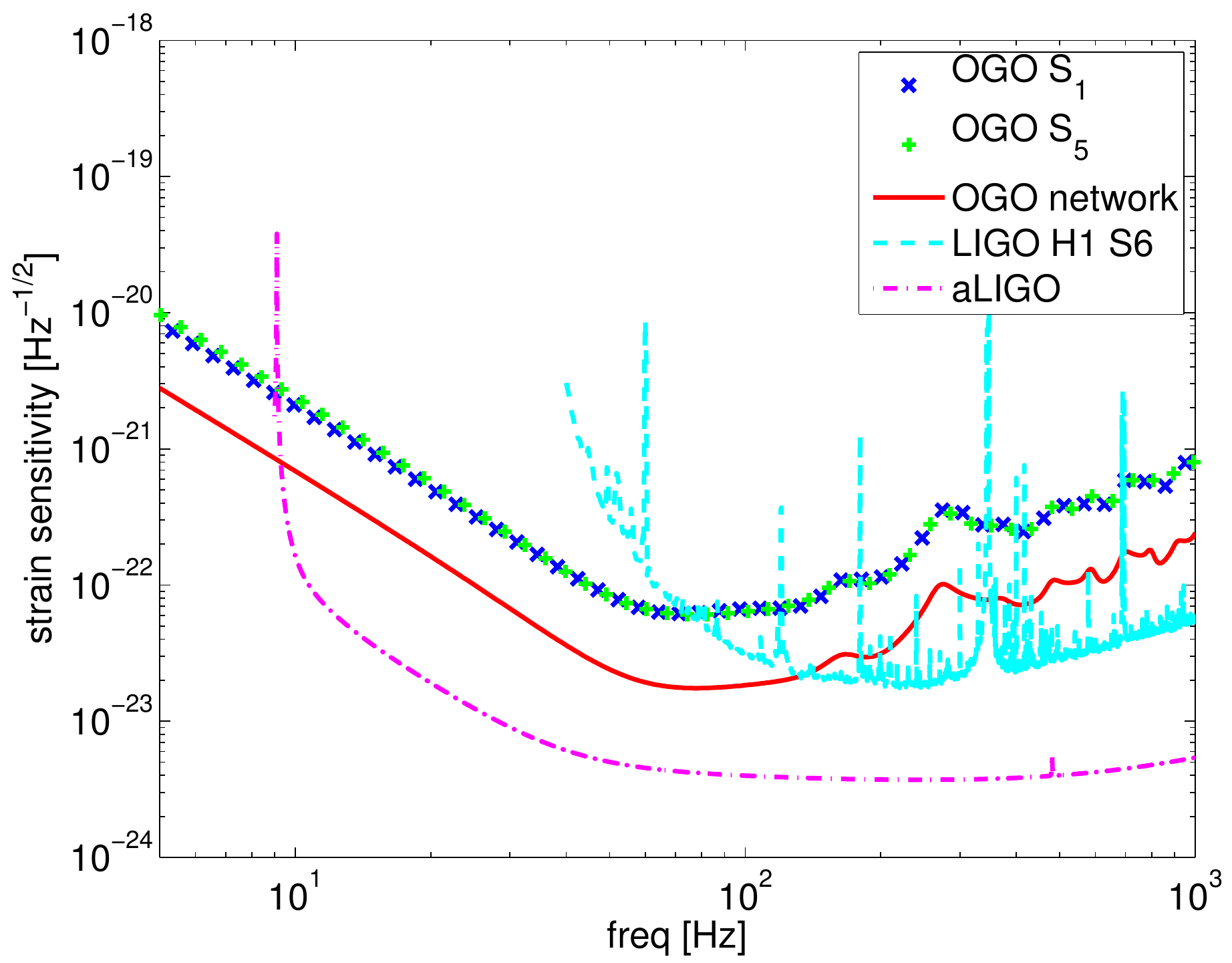}\\
  \caption{Sensitivities for two single DFI combinations ($S_1$, \emph{blue crosses} and $S_5$, \emph{green plus signs}) of OGO (with $L\approx1400$\,km) and for the full OGO network sensitivity (scaled from $S_5$, \emph{red solid line}). For comparison, the dashed lines show sensitivities for initial LIGO (H1 during science run S6, from Ref.~\cite{Abadie2010a}, \emph{cyan dashed line}) and aLIGO (design sensitivity for high-power, zero detuning configuration, from Ref.~\cite{aLIGO_sensitivity}, \emph{magenta dash-dotted line}).}
  \label{F:Sens}
 \end{figure}

 Therefore, we simply approximate the network sensitivity of the full detector as $\sqrt{\widetilde{S}_{h, {\rm net}}} = \sqrt{\widetilde{S}_{h,5}/12}$.
 Note that the potential loss of some links would imply that not all generators can be formed. We can lose up to 6 links and still be able to form a DFI stream (but probably  only one). The number of lost links (and which links are lost exactly) will affect the network sensitivity. In our estimations below we deal with the idealized situation and
 assume that no links are lost.

 We plot the sensitivity curves for individual combinations and the network sensitivity in Fig.~\ref{F:Sens}.
 For comparison we also show the design sensitivity curves of initial LIGO (S6 science run~\cite{Abadie2010a}) and advanced LIGO (high laser power configuration with zero detuning of the signal recycling mirror~\cite{aLIGO_sensitivity}).
 Indeed one can see that the sensitivities of the individual OGO configurations are similar to each other and close to initial LIGO.
 The network sensitivity of OGO lies between LIGO and aLIGO sensitivities.
 OGO as expected outperforms aLIGO below 10\,Hz, where the seismic noise on the ground becomes strongly dominant.

 \subsection{General performance of the DFI scheme}
 \label{sec:performance}

 \begin{figure}[th]
  \includegraphics[width=\columnwidth, keepaspectratio=true]{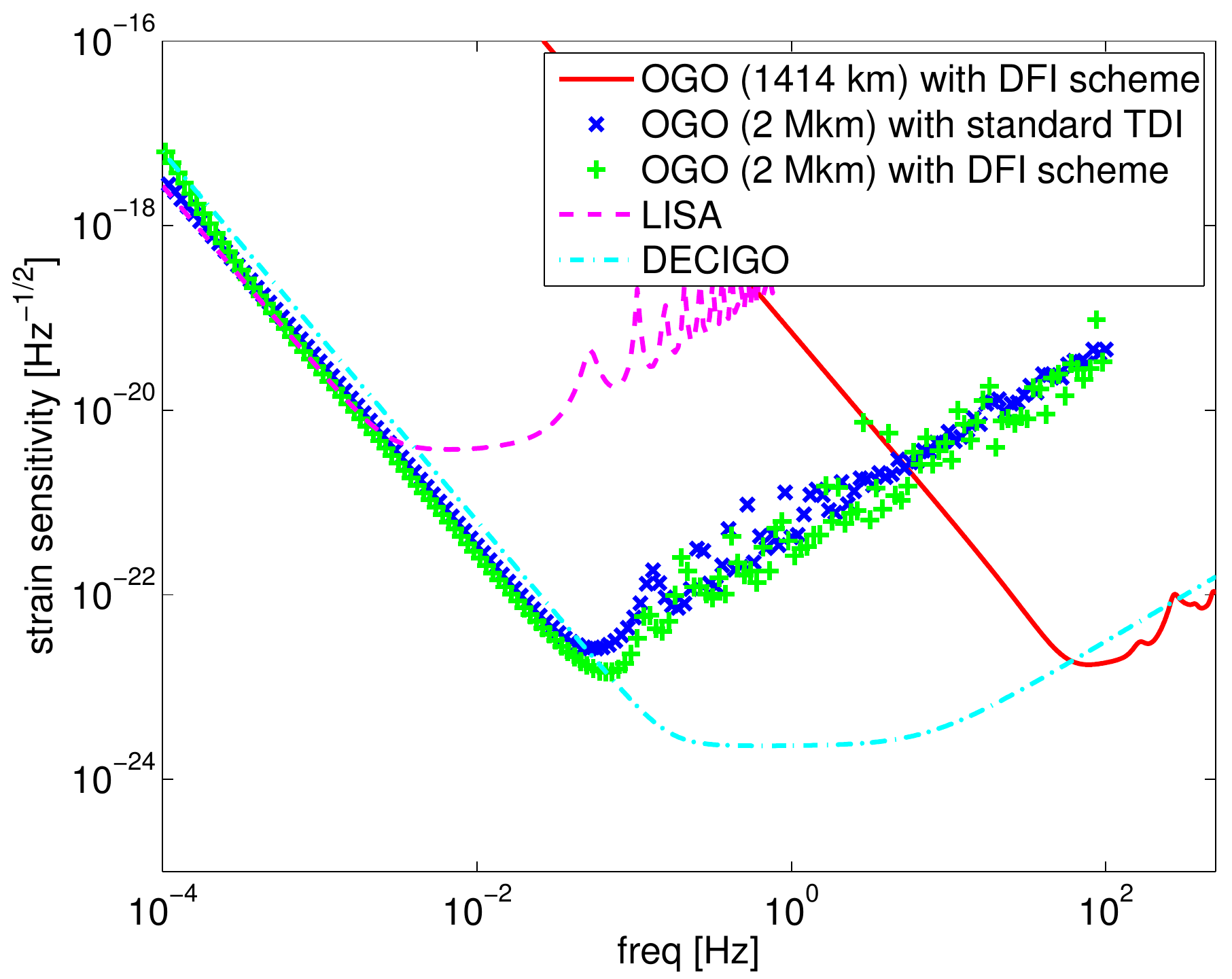}\\
  \caption{Network sensitivities, scaled from $S_5$, of standard OGO (with DFI, arm length 1414 km, \emph{red solid line}) compared to an OGO-like detector with spacecraft separation of $2\cdot10^9$ m, with either full DFI scheme (\emph{blue crosses}) or standard TDI only (\emph{green plus signs}). Also shown for comparison are (classic) LISA ($5\cdot10^9$ m, network sensitivity, \emph{magenta dashed line}, from Ref.~\cite{Larson2000}) and DECIGO (using the fitting formula Eq.~(20) from Ref.~\cite{Yagi2013}, \emph{cyan dash-dotted line}).}
  \label{F:ogo_TDI_comparison_2Mkm}
 \end{figure}

 Having derived the full sensitivity curve of the OGO mission design with $L\approx1400$\,km as an exemplary implementation of the three-dimensional DFI scheme in space, let us take a step back and analyze the general performance of a DFI-enabled detector.
 These features are also what led us to consider the octahedron configuration in the first place.

 Specifically, let us look in more detail at the low frequency asymptotic behavior of the transfer functions and sensitivity curves.
 We consider a LISA-like configuration with two laser noise free combinations: an unequal arm Michelson (\mbox{TDI-$X$}) and a Sagnac combination (\mbox{TDI-$\alpha$}). 
 Let us assume for a moment that the only noise source is shot noise, which at low frequencies ($\epsilon \ll 1$) scales as $\sqrt{\widetilde{S}_{\mathrm{n},X}} \sim \epsilon^2$ and $\sqrt{\widetilde{S}_{\mathrm{n},\alpha}} \sim \epsilon^1$ for those two combinations, respectively.

 The GW transfer function, for both TDI combinations, scales as $\mathcal{T}_\alpha,\mathcal{T}_X \sim \epsilon^2$; therefore, the sensitivity curves scale as $\sqrt{\widetilde{S}_{h,\alpha}} \sim \sqrt{\widetilde{S}_{\mathrm{n},\alpha}}/\mathcal{T}_\alpha \sim \epsilon^{-1}$ for \mbox{TDI-$\alpha$} and $\sqrt{\widetilde{S}_{h,X}} \sim \sqrt{\widetilde{S}_{\mathrm{n},X}}/\mathcal{T}_X \sim \epsilon^{0}$ for \mbox{TDI-$X$}.
 We see that a LISA-like \mbox{TDI-$X$}-combination has a \emph{flat} shot-noise spectrum at low frequencies, corresponding to a flat total detector sensitivity if all other dominant noise sources can be canceled -- which looks extremely attractive.

 Thus, a naive analysis suggests that the acceleration and laser noise free combinations for an octahedron detector could yield a flat sensitivity curve at low frequencies.
 Checking this preliminary result with a more careful analysis was the main motivation for the research presented in this article.

 In fact, as we have seen in Sec.~\ref{sec:transfer_function}, the full derivation delivers transfer functions that, in leading order of $\epsilon$, go as $\mathcal{T}_{1,2,\ldots,7} \sim \epsilon^3$.
 This implies that the sensitivity for laser and acceleration noise free combinations behaves as $\sqrt{\widetilde{S}_{h,1,2,\ldots,7}}/\mathcal{T}_{1,2,\ldots,,7} \sim \epsilon^{-2}$, which is similar to the behavior of acceleration noise.
 In other words, the combinations eliminating the acceleration noise also cancel a significant part of the GW signal at low frequencies.

 In fact, we find that a standard LISA-like TDI-enabled detector of the same arm length and optical configuration as OGO could achieve a similar low-frequency sensitivity (at few to tens of Hz) with an acceleration noise requirement of only $\sim10^{-12}$~m/s${}^2\,\sqrt{\mathrm{Hz}}$.
 This assumes negligible spacecraft jitter and that no other noise sources (phase-meter noise, sideband noise, thermal noise) limit the sensitivity, which at this frequency band would behave differently than in the LISA band.
 In fact, the GOCE mission~\cite{Drinkwater2003} has already demonstrated such acceleration noise levels at mHz frequencies~\cite{Sechi2011}, and therefore this seems a rather modest requirement at OGO frequencies.
 We therefore see that such a short-arm-length OGO would actually only be a more complicated alternative to other feasible mission designs.

 In addition, it is hard to see from just the comparison with ground-based detectors in Fig.~\ref{F:Sens} how exactly the DFI method itself influences the final noise curve of OGO, and how much of its shape is instead determined by the geometrical and technical parameters of the mission concept (arm length, laser power, telescope size).
 Also, the secondary technological noise sources of a space mission in the comparatively high-frequency band of this exemplary OGO implementation are somewhat different from more well-studied missions like LISA and DECIGO.

 Therefore, to disentangle these effects, we will now tentatively study a different version of OGO based on the alternative orbit with an average arm length of $2\cdot10^9$\,m, as mentioned in Sec.~\ref{S:Orbit}.
 It requires further study to determine whether a stable octahedron constellation and the DFI scheme are possible on such an orbit, but assuming they are, we can compute its sensitivity as before.

 In Fig.~\ref{F:ogo_TDI_comparison_2Mkm}, we then compare this longer-baseline DFI detector with another detector with the same geometry and optical components, but without the DFI technique, using instead conventional TDI measurements.
 Here, we are in a similar frequency range as LISA and therefore assume similar values for the acceleration noise of $3 \cdot 10^{-15}$~m/s${}^2\,\sqrt{\mathrm{Hz}}$~\cite{LisaYellow}
 and secondary noise sources (phase meter, thermal noise, etc.; see Sec.~\ref{sec:feasibility}).

 However, there is another noise source, spacecraft jitter, which is considered subdominant for LISA, but might become relevant for both the TDI and DFI versions of the $2\cdot10^9$\,m OGO-like detector.
 Jitter corresponds to the rotational degrees of freedom between spacecraft, and its coupling into measurement noise is not fully understood.
 We have therefore computed both sensitivities without any jitter.
 It seems possible that at least the part of jitter that couples linearly into displacement noise could also be canceled by DFI, or that an extension of DFI (e.g. more links) could take better care of this,
 and therefore that the full OGO with DFI would look more favorable compared to the TDI version when nonvanishing jitter is taken into account.

 Generally, as one goes for longer arm lengths, the DFI scheme will perform better in comparison to the TDI scheme.
 At the high-frequency end of the sensitivity curves, both schemes are limited by shot noise and the respective GW transfer functions.
 Since the shot-noise level does not depend on the arm length, it remains the same for all relevant frequencies.
 Therefore, as the arm length increases, the high-frequency part of the sensitivity curves moves to the low-frequency regime in parallel (i.e. the corner frequency of the transfer function is proportional to $1/L$).
 This is the same for both schemes.

 On the other hand, in the low-frequency regime of the sensitivity curves the two schemes perform very differently.
 For TDI, the low-frequency behavior is limited by acceleration noise, while for DFI this part is again limited by shot noise and the GW transfer function.
 When the arm length increases, the low-frequency part of the sensitivity curve in the TDI scheme moves to lower frequencies in proportion to $1/\sqrt{L}$; while for DFI, it moves in proportion to $1/L$.

 Graphically, when the arm length increases, the high-frequency parts of the sensitivity curves in both schemes move toward the lower-frequency regime in parallel, while the low-frequency part of the sensitivity curve for DFI moves faster than for TDI.

 Under the assumptions given above, we find that an arm length of $2\cdot10^9$\,m is close to the transition point where the sensitivities of TDI and DFI are almost equal, as shown in Fig.~\ref{F:ogo_TDI_comparison_2Mkm}.
 At even longer arm lengths, employing DFI would become clearly advantageous.

 Of course, these considerations show that a longer-baseline detector with good sensitivity in the standard space-based detector frequency band of interest would make a scientifically much more interesting case than the default short-arm OGO which we presented first.
 However, as no study on the required orbits has been done so far, we consider such a detector variant to be highly hypothetical and not worthy of a detailed study of technological feasibility and scientific potential yet.
 Instead, for the remainder of this paper, we concentrate again on the conservative 1400\,km version of OGO.
 Although the sensitivity curve in Fig.~\ref{F:Sens} already demonstrates its limited potential, we will attempt to neutrally assess its advantages, limitations and scientific reach.

 \subsection{Technological feasibility}
 \label{sec:feasibility}
 Employing DFI requires a large number of spacecraft but on the other hand allows us to relax many of the very strict technological requirements of other space-based GW detector proposals such as (e)LISA and DECIGO.
 Specifically, the clock noise is canceled by design, so there is no need for a complicated clock tone transfer chain~\cite{barke2010}.
 Furthermore, OGO does not require a drag-free technology, and the configuration has to be stabilized only as much as required for the equal arm length assumption to hold.
 This strongly reduces the requirements on the spacecraft thrusters.
 Also, for the end mirrors, which have to be mounted on the same monolithic structure for all four laser links per spacecraft, it is not required that they are freefalling.
 Instead, they can be fixed to the spacecraft.

 Still, to reach the shot-noise-only limited sensitivity shown in Fig.~\ref{F:Sens}, the secondary noise contributions from  all components of the measurement system must be significantly below the shot-noise level.
 Considering a shot-noise level of about $2\cdot 10^{-17}\,\mathrm{m}/\sqrt{\mathrm{Hz}}$ -- which is in agreement with the value derived earlier for the 1400\,km version of OGO -- this might be challenging.

 When actively controlling the spacecraft position and hence stabilizing the distance and relative velocity between the spacecraft, we will be able to lower the heterodyne frequency of the laser beat notes drastically.
 Where LISA will have a beat note frequency in the tens of MHz, with OGO's short arm length we could be speaking of kHz or less and might even consider a homodyne detection scheme as in LIGO.
 This might in the end enable us to build a phase meter capable of detecting relative distance fluctuations with a sensitivity of $10^{-17}\,\mathrm{m}/\sqrt{\mathrm{Hz}}$ or below as required by OGO.

 As mentioned before, temperature noise might be a relevant noise source for OGO:
 The relative distance fluctuations on the optical benches due to temperature fluctuations and the test mass thermal noise must be significantly reduced in comparison to LISA.
 But even though the LISA constellation is set in an environment which is naturally more temperature stable, stabilization should be easier for the higher-frequency OGO measurement band.
 A requirement of $10^{-17}\,\mathrm{m}/\sqrt{\mathrm{Hz}}$ could be reached by actively stabilizing the temperature down to values of 1\,nK/$\sqrt{\mathrm{Hz}}$ at the corner frequency.

 Assuming future technological progress, optimization of the optical bench layout could also contribute to mitigating this constraint, as could the invention of thermally more stable materials for the optical bench.
 Most likely, this challenge can be solved only with a combination of the mentioned approaches.

 The same is true for the optical path length stability of the telescopes.
 We estimate the required pointing stability to be roughly similar to the LISA mission requirements.

\subsection{Shot-noise reduction}
 \label{sec:shot_noise_reduction}

 Assuming the requirements from the previous section can be met, the timing and acceleration noise free combinations of the OGO detector are dominated by shot noise, and any means of reducing the shot noise will lead to a sensitivity improvement over all frequencies.
 In this subsection, we discuss possible ways to achieve such a reduction.

 The most obvious solution is to increase laser power, with an achievable sensitivity improvement that scales with $\sqrt{P}$.
 However, the available laser power is limited by the power supplies available on a spacecraft.
 Stronger lasers are also heavier and take more place, making the launch of the mission more difficult.
 Therefore, there is a limit to simply increasing laser power, and we want to shortly discuss more advanced methods of shot-noise reduction.

 One such hypothetical possibility is to build cavities along the links between spacecraft, similar to the DECIGO design~\cite{Ando2010}.
 The shot noise would be decreased due to an increase of the effective power stored in the cavity.
 Effectively, this also results in an increase of the arm length.
 Note, however, that the sensitivity of OGO with cavities cannot simply be computed by inserting effective power and arm length into our previously derived equations.
 Instead, a rederivation of the full transfer function along the lines of Ref.~\cite{Rakhmanov2005} is necessary.

 Alternatively, squeezed light~\cite{Schnabel10} is a way to directly reduce the quantum measurement noise, which has already been demonstrated in ground-based detectors~\cite{SqueezedGEO,Khalaidovski12}.
 However, squeezing in a space-based detector is challenging in many aspects due to the very sensitive procedure and would require further development.

\section{Scientific perspectives}
 \label{S:Sources}
 In this section, we will discuss the science case for our octahedral GW detector (with an arm length of 1400\,km) by considering the most important potential astrophysical sources in its band of sensitivity.
 Using the full network sensitivity, as derived above, the best performance of OGO is at 78\,Hz, between the best achieved performance of initial LIGO during its S6 science run and the anticipated sensitivity for advanced LIGO.
 OGO outperforms the advanced ground-based detectors below 10\,Hz, where the seismic noise strongly dominates.
 In this analysis, we will therefore consider sources emitting GWs with frequencies between 1\,Hz and 1\,kHz, concentrating on the low end of this range.

 Basically, those are the same sources as for ground-based detectors, which include compact binaries coalescences (CBCs), asymmetric single neutron stars (continuous waves, CWs), binaries containing intermediate-mass black holes (IMBHs), burst sources (unmodeled short-duration transient signals), and a cosmological stochastic background.

 We will go briefly through each class of sources and consider perspectives of their detection.
 As was to be expected from the sensitivity curve in Fig.~\ref{F:Sens}, in most categories OGO performs better than initial ground-based detectors, but does not even reach the potential of the advanced generation currently under commissioning.

 Therefore, this section should be understood not as an endorsement of actually building and flying an OGO-like mission, but just as an assessment of its (limited, but existing) potentials.
 This demonstrates that an octahedral GW detector employing DFI in space is in principle capable of scientifically interesting observations, even though improving its performance to actually surpass existing detectors or more mature mission proposals still remains a subject of further study.

 In addition, we put a special focus on areas where OGO's design offers some specific advantages.
 These include the triangulation of CBCs through joint detection with ground-based detectors as well as searching for a stochastic GW background and for additional GW modes.

 Note that the hypothetical $2\cdot10^9$\,m variant of OGO (see Secs. \ref{S:Orbit} and \ref{sec:performance}) would have a very different target population of astrophysical sources due to its sensitivity shift to lower frequencies.
 Such a detector would still be sensitive to CBCs, IMBHs, and stochastic backgrounds, probably much more so.
 But instead of high-frequency sources like CW pulsars and supernova bursts, it would start targeting supermassive black holes, investigating the merging history of galaxies over cosmological scales.

 However, as this detector concept relies on an orbit hypothesis not studied in any detail, we do not consider it mature enough to warrant a study of potential detection rates in any detail, and we therefore only refer to established reviews of the astrophysical potential in the frequency band of LISA and DECIGO, e.g. Ref.~\cite{Sathya2009}.

 \subsection{Coalescing compact binaries}
 Heavy stars in binary systems will end up as compact objects (such as NSs or BHs) inspiralling around each other, losing orbital energy and angular momentum through gravitational radiation.
 Depending on the proximity of the source and the detector's sensitivity, we could detect GWs from such a system a few seconds up to a day before the merger and the formation of a single spinning object.
 These CBCs are expected to be the strongest sources of GWs in the frequency band of current GW detectors.

 To estimate the event rates for various binary systems, we will follow the calculations outlined in Ref.~\cite{Abadie2010b}.
 To compare with predictions for initial and advanced LIGO (presented in Ref.~\cite{Abadie2010b}), we also use only the inspiral part of the coalescence to estimate the \emph{horizon distance} (the maximum distance to which we can observe a given system with a given signal-to-noise ratio (SNR)).
 We use here the same detection threshold on signal-to-noise ratio, a SNR of $\rho=8$, as in Ref.~\cite{Abadie2010b} and consider the same fiducial binary systems: NS-NS (with $1.4 \msun$ each), BH-NS (BH mass $10 \msun$, NS with $1.4 \msun$), and BH-BH ($10 \msun$ each).

 For a binary of given masses, the sky-averaged horizon distance is given by
 \begin{equation}
  D_{h} = \frac{4 \sqrt{5} \, G^{\frac{5}{6}} \, \mu^{\frac{1}{2}} \, M^{\frac{1}{3}}}{\sqrt{96} \, \pi^{\frac{2}{3}} \, c^{\frac{3}{2}} \, \rho}
 \sqrt{\int_{f_{\rm min}}^{f_{\rm ISCO}} \frac{f^{-\frac{7}{3}}}{\widetilde{S}_{\rm h}(f)}\,\mathrm{d}f} \; .
 \label{eq:Dhorizon}
 \end{equation}
 Here, $M=M_1+M_2$ is the total mass and $\mu={M_1M_2}/{M}$ is the reduced mass of the system.
 We have used a lower cutoff of $f_{\rm min} = 1$\,Hz, and at the upper end the frequency of the innermost stable circular orbit is $f_{\rm ISCO} = c^3/(6^{3/2}\pi\ G\ M)$ Hz, which conventionally is taken as the end of the inspiral.

 Now, for any given type of binary (as characterized by the component masses), we obtain
 the observed event rate (per year) using  $\dot{N}=R \cdot N_{\mathrm{G}}$, where we have adopted the approximation for the number of galaxies inside the visible volume from Eq.~(5) of Ref.~\cite{Abadie2010b}:
 \begin{equation}\label{NG}
  N_{\mathrm{G}}=\frac{4}{3}\pi \left(\frac{D_{\mathrm{h}}}{\mathrm{Mpc}}\right)^3 (2.26)^{-3} \cdot 0.0116 \, ,
 \end{equation}
 and the intrinsic coalescence rates $R$ per Milky-Way-type galaxy are given in Table 2 of Ref.~\cite{Abadie2010b}.

 A single DFI combination $S_i$ has annual rates similar to initial LIGO, and the results for the network sensitivity of full OGO are summarized in Table~\ref{T:cbc_rates}.
 For each binary, we give three numbers following the uncertainties in the intrinsic event rate (``pessimistic'', ``realistic'', ``optimistic'') as introduced in Ref.~\cite{Abadie2010b}.

 \begin{table}
  \begin{tabular}{l l l l}
   \hline\hline
   & NS-NS & NS-BH & BH-BH \\
   \hline
   OGO           & (0.002,   0.2,    2.2)    & (0.001, 0.06,  2.0)      & (0.003, 0.1,    9) \\
   LIGO          & (2e-4, 0.02,   0.2)    & (7e-5, 0.004, 0.1)    & (2e-4,  0.007,  0.5) \\
   aLIGO         & (0.4,  40,     400)    & (0.2,  10,    300)    & (0.4,   20,     1000) \\
   \hline\hline
  \end{tabular}
  \caption{Estimated yearly detection rates for CBC events, given in triplets of the form (lower limit, realistic value, upper limit) as defined in Ref.~\cite{Abadie2010b}.}
  \label{T:cbc_rates}
 \end{table}

 From this, we see that OGO achieves detection rates an order of magnitude better than initial LIGO.
 But we still expect to have only one event in about three years of observation assuming ``realistic''
 intrinsic coalescence rates.
 The sensitivity of aLIGO is much better than for OGO above 10 Hz, and the absence of seismic noise does not help OGO much because the absolute sensitivities below 10 Hz are quite poor and only a very small fraction of SNR is contributed from the lower frequencies.
 This is the reason why OGO cannot compete directly with aLIGO in terms of total CBC detection rates, which are about two orders of magnitude lower.

 However, OGO does present an interesting scientific opportunity when run in parallel with aLIGO.
 If OGO indeed detects a few events over its mission lifetime, as the realistic predictions allow, it can give a
 very large improvement to the sky localization of these sources.
 Parameter estimation by aLIGO alone typically cannot localize signals enough for efficient electromagnetic follow-up
 identification.
 However, in a joint detection by OGO and aLIGO, triangulation over the long baseline between space-based OGO and ground-based aLIGO would yield a fantastic angular resolution.
 As signals found by OGO are very likely to be picked up by aLIGO as well, such joint detections indeed seem promising.
 Additionally, the three-dimensional configuration and independent channels of OGO potentially allow a more accurate parameter estimation than a network of two or three simple L-shaped interferometers could achieve.

 \subsection{Stochastic background}

 There are mainly two kinds of stochastic GW backgrounds~\cite{Allen99, Maggiore00}:
 The first is the astrophysical background (sometimes also called astrophysical foreground), arising from unresolved astrophysical sources such as compact binaries~\cite{Farmer03} and core-collapse supernovae~\cite{Ferrari99}.
 It provides important statistical information about distribution of the sources and their parameters.
 The second is the cosmological background which was generated by various mechanisms in the early Universe~\cite{Brustein95, Turner97,Ananda07}.
 It carries unique information about the very beginning of the Universe ($\sim 10^{-28}$\,s).
 Thus, the detection of the GW stochastic background is of great interest.

 Currently, there are two ways to detect the stochastic GW background.
 One of them~\cite{Hogan01} takes advantage of the null stream (e.g. the Sagnac combination of LISA).
 By definition, the null stream is insensitive to gravitational radiation, while it suffers from the same noise sources as the normal data stream.
 A comparison of the energy contained in the null stream and the normal data stream allows us to determine whether the GW stochastic background is present or not.
 The other way of detection is by cross-correlation~\cite{Allen99,Seto06} of measurements taken by different detectors.
 In our language, this uses the GW background signal measured by one channel as the template for the other channel.
 In this sense, the cross-correlation can be viewed as matched filtering.
 Both ways require redundancy, i.e.\ more than one channel observing the same GW signal with independent noise.

 Luckily, the octahedron detector has plenty of redundancy, which potentially allows precise background detection.
 There are in total 12 dual-way laser links between spacecraft, forming 8 LISA-like triangular constellations.
 Any pair of two such LISA-like triangles that does not share common links can be used as an independent correlation.
 There are 16 such pairs within the octahedron detector.
 Within each pair, we can correlate the orthogonal TDI variables A, E and T (as they are denoted in LISA~\cite{TintoDhurandhar}).
 Altogether, there are $16\times3^2=144$ cross-correlations.

 And we have yet more information encoded by the detector, which we can access by considering that any two connected links form a Michelson interferometer, thus providing a Michelson-TDI variable.
 Any two of these variables that do not share common links can be correlated.
 There are in total 36 such variables, forming 450 cross-correlations, from which we can construct the optimal total sensitivity.

 Furthermore, each of these is sensitive to a different direction on the sky.
 So the octahedron detector has the potential to detect anisotropy of the stochastic background.
 However, describing an approach for the detection of anisotropy is beyond the scope of this feasibility study.

 Instead, we will present here only an order of magnitude estimation of the total cross-correlation SNR.
 Usually, it can be expressed as
 \begin{equation}
  \mathrm{SNR} = \frac{3H_0^2}{10\pi^2}\sqrt{T_{\mathrm{obs}}}\left[ 2 \sum_{k,l}\int_0^\infty {\rm d}f\frac{\gamma_{kl}^2(f)\Omega_{\rm gw}^2(f)} {f^6\widetilde{S}_{{\rm h}, k}(f)\widetilde{S}_{{\rm h}, l}(f)} \right]^{\frac{1}{2}}\,,
 \end{equation}
 where $T_{\mathrm{obs}}$ is the observation time, $\Omega_{\rm gw}$ is the fractional energy-density of the Universe in a GW background, $H_0$ the \emph{Hubble constant}, and $\widetilde{S}_{{\rm h}, k}(f)$ is the effective sensitivity of the $k$-th channel.
 $\gamma_{kl}(f)$ denotes the \emph{overlap reduction function} between the $k$-th and $l$-th channels, introduced by Flanagan~\cite{Flanagan93}.
 \begin{equation}
  \gamma_{kl}(f) = \frac{5}{8\pi}\sum_{p=+,\times}\int {\rm d}\hat{\Omega}\,\mathrm{e}^{2\pi {\rm i} f \hat{\Omega}\cdot \Delta \mathbf{x}/c}F_k^p(\hat{\Omega})F_l^p(\hat{\Omega})\, ,
 \end{equation}
 where $F_k^p(\hat{\Omega})$ is the antenna pattern function.
 As mentioned in the previous section, there might be 12 independent DFI solutions.
 These DFI solutions can form $12\times 11/2 = 66$ cross-correlations.
 According to Ref.~\cite{Allen99}, we know $\gamma_{kl}^2(f)$ varies between $0$ and $1$.
 As a rough estimate, we approximate $\sum_{k,l}\gamma_{kl}^2(f)\sim 10$; hence, we get the following result for OGO:
 \begin{equation}
  {\rm SNR} = 2.57\left(\frac{H_0}{72\,\frac{{\rm km} \, {\rm s^{-1}}}{{\rm Mpc}}}\right)^2\left(\frac{\Omega_{\rm gw}}{10^{-9}}\right)\left(\frac{T_{\mathrm{obs}}}{10\,{\rm yr}}\right)^{\frac{1}{2}}\,.
 \end{equation}
 Initial LIGO has set an upper limit of $6.9\cdot 10^{-6}$ on $\Omega_{\rm gw}$~\cite{nature09}, and aLIGO will be able to detect the stochastic background at the $1\cdot 10^{-9}$ level~\cite{nature09}.
 Hence, our naive estimate of OGO's sensitivity to the GW stochastic background is similar to that of aLIGO.
 Actually, an optimal combination of all the previously-mentioned possible cross-correlations would potentially result in an even better detection ability for OGO.

 \subsection{Testing alternative theories of gravity}

 \begin{figure}[th]
  \includegraphics[width=\columnwidth, keepaspectratio=true]{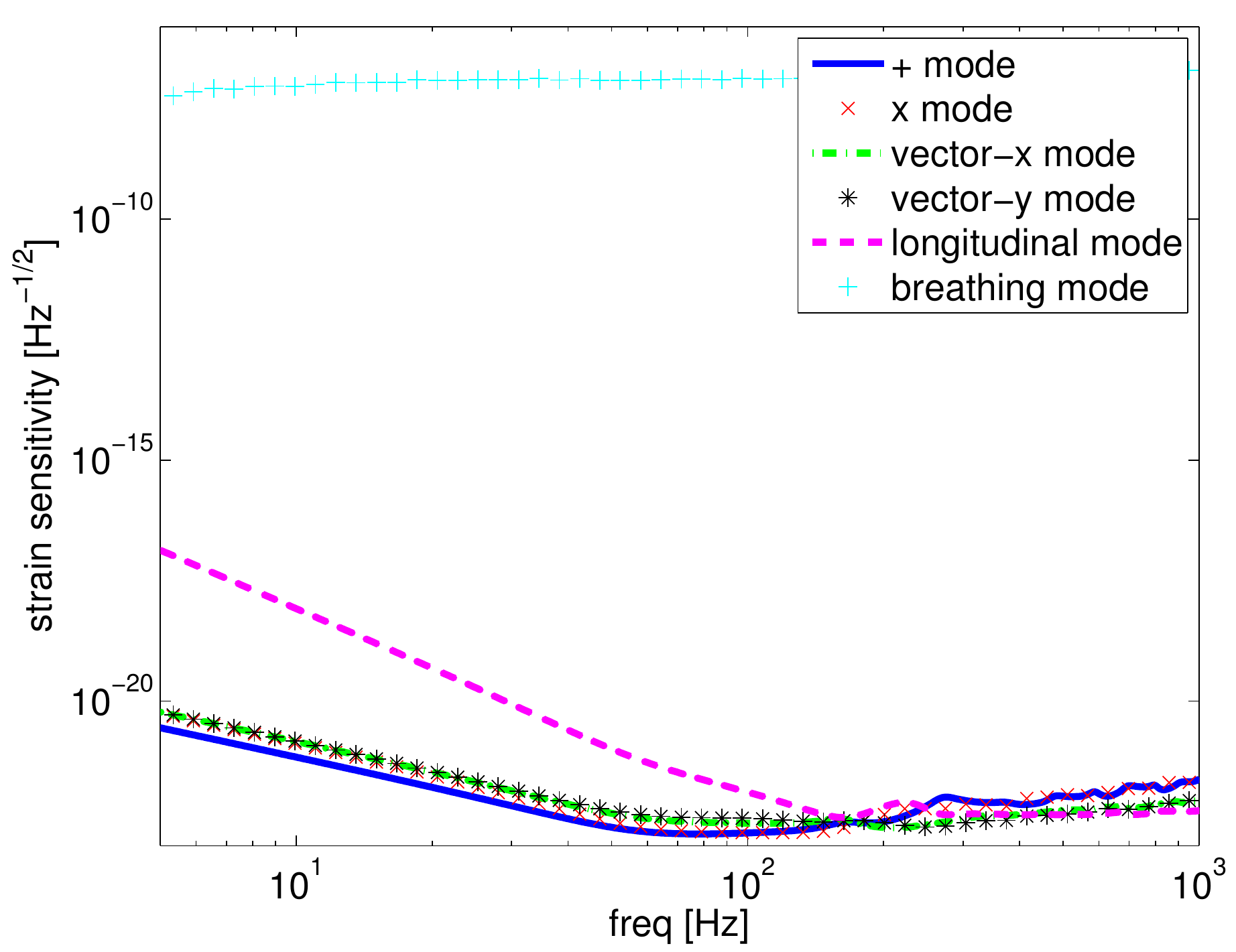}
  \caption{Relative sensitivity of the full OGO network (scaled from S5 combination) to alternative polarizations:
           $+$ mode (\emph{blue solid line}),
           $x$ mode (\emph{red crosses}),
           vector-$x$ mode (\emph{green dash-dotted line}),
           vector-$y$ mode (\emph{black stars}),
           longitudinal mode (\emph{magenta dashed line}),
           and breathing mode (\emph{cyan plus signs}).
          }
  \label{F:altern_polar}
 \end{figure}

 In this section we will consider OGO's ability to test predictions of General Relativity against alternative theories.
 In particular, we will estimate the sensitivity of the proposed detector to all six polarization modes that could be present in (alternative) metric theories of gravitation~\cite{Hohmann2012}.
 We refer to Ref.~\cite{Eardley1973} for a discussion on polarization states, which are (i) two transverse-traceless (tensorial) polarizations usually denoted as $+$ and $\times$, (ii) two scalar modes called breathing (or common) and longitudinal and (iii) two vectorial modes.
 We also refer to Refs.~\cite{Will_LR,Gair_LR} for reviews on alternative theories of gravity.

 We have followed the procedure for computing the sensitivity of OGO, as outlined above, for the four modes not present in General Relativity, and we compare those sensitivities to the results for the $+,  \times$ modes as presented in Fig.~\ref{F:Sens}.
 The generalization of the transfer function used in this paper [Eq.~\ref{Eq:TF}] for other polarization modes is given in Ref.~\cite{Chamberlin_2012}.

 We have found that all seven generators show similar sensitivity for each mode.
 OGO is not sensitive to the common (breathing) mode, which is not surprising as it can be attributed to a common displacement noise, which we have removed by our procedure.
 The sensitivity to the second (longitudinal) scalar mode scales as $\epsilon^{-4}$ at low frequencies and is much worse than the sensitivity to the $+, \times$ polarizations below 200\,Hz.
 However, OGO is more sensitive to the longitudinal mode (by about an order of magnitude) above 500\,Hz.
 The sensitivity of OGO to vectorial modes is overall similar to the $+, \times$ modes: it is by a few factors less sensitive to vectorial modes below 200\,Hz and by similar factors more sensitive above 300\,Hz.
 These sensitivities are shown in Fig.~\ref{F:altern_polar}.

 \subsection{Pulsars -- Continuous Waves}

 CWs are expected from spinning neutron stars with nonaxisymmetric deformations.
 Spinning NSs are already observed as radio and gamma-ray pulsars.
 Since CW emission is powered by the spindown of the pulsar, the strongest emitters are the pulsars with high spindowns, which usually are young pulsars at rather high frequencies.
 Note that the standard emission model~\cite{Jaranowski1998} predicts a gravitational wave frequency $f_{\mathrm{gw}}=2f$, while alternative models like free precession~\cite{Jones2001} and $r$-modes~\cite{Andersson1998} also allow emission at $f_{\mathrm{gw}}=f$ and $f_{\mathrm{gw}}=\tfrac{4}{3}f$, where $f$ is the NS spin frequency.

\begin{figure}[th]
  \centering
  \includegraphics[width=\columnwidth]{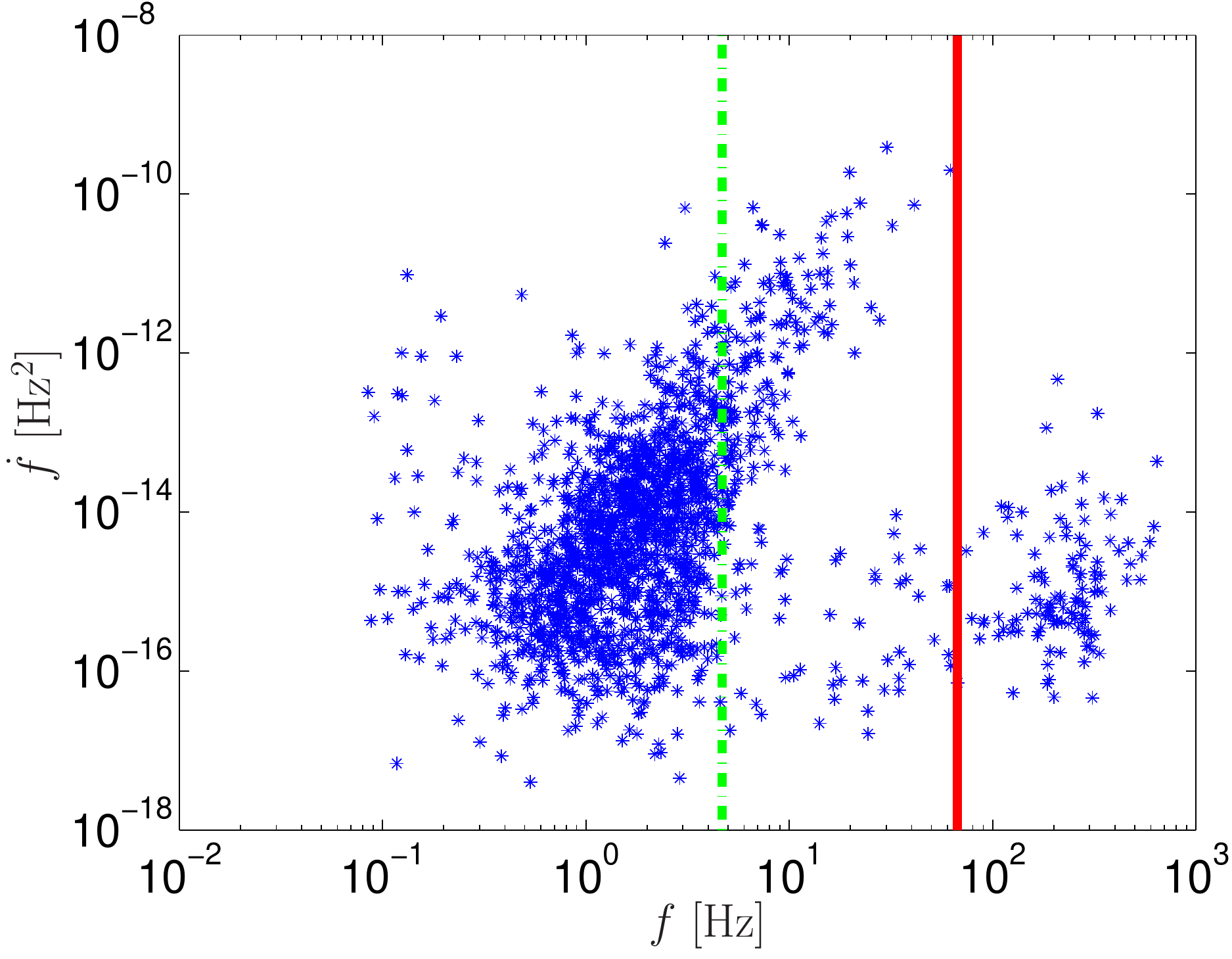}
  \caption{Population of currently known pulsars in the frequency-spindown plain ($f$-$\dot{f}$).
           OGO could beat initial LIGO left of the red solid line and Advanced LIGO left of the green dashed line.
           Data for this plot were taken from Ref.~\cite{ATNF} on March 2, 2012.}
  \label{fig:pulsars_atnf}
 \end{figure}

 OGO has better sensitivity than initial LIGO below 133\,Hz, has its best sensitivity around 78\,Hz, and is better than aLIGO below 9\,Hz. 
 This actually fits well with the current radio census of the galactic pulsar population, as given by the ATNF catalog~\cite{ATNF}.
 As shown in Fig.~\ref{fig:pulsars_atnf}, the bulk of the population is below $\sim$ 10\,Hz, and also contains many low-frequency pulsars with decent spindown values, even including a few down to $\sim$ 0.1\,Hz.

 We estimate the abilities of OGO to detect CW emission from known pulsars following the procedure outlined in Ref.~\cite{Abadie2011} for analysis of the Vela pulsar.
 The GW strain for a source at distance $D$ is given as
 \begin{equation}
  h_0 = \frac{4 \pi^2 G  I_{zz} \epsilon f^2}{c^4 D} \, ,
 \end{equation}
 where  $\epsilon$ is the ellipticity of the neutron star and we assume a canonical momentum of inertia $I_{zz} = 10^{38}$\,kg\,m$^2$.
 After an observation time $T_{\rm obs}$, we could detect a strain amplitude
 \begin{equation}
  h_0 = \Theta\sqrt{\frac{S_\mathrm{h}}{T_{\mathrm{obs}}}} \, .
 \end{equation}
 The statistical factor is $\Theta\approx11.4$ for a fully coherent targeted search with the canonical values of 1\,\% and 10\,\% for false alarm and false dismissal probabilities, respectively~\cite{Abbott2004}.
 We find that, for the Vela pulsar (at a distance of 290\,pc and a frequency of $f_{\mathrm{Vela,gw}}=2 \cdot 11.19$\,Hz), with $T_{\mathrm{obs}}=30$ days of observation, we could probe ellipticities as low as $\epsilon \sim 5\cdot10^{-4}$ with the network OGO configuration.
 Several known low-frequency pulsars outside the aLIGO band would also be promising objectives for OGO targeted searches.

 All-sky searches for unknown pulsars with OGO would focus on the low-frequency range not accessible to aLIGO with a search setup comparable to current Einstein@Home LIGO searches~\cite{Aasi2013}.
 As seen above, the sensitivity estimate factors into a search setup related part $\Theta / \sqrt{T_{\mathrm{obs}}}$ and the sensitivity $\sqrt{S_\mathrm{h}}$.
 Therefore, scaling a search with parameters identical to the Einstein@Home S5 runs to OGO's best sensitivity at 76\,Hz would reach a sensitivity of $h_0 \approx 3 \cdot 10^{-25}$.
 This would, for example, correspond to a neutron star ellipticity of $\epsilon \sim 4.9 \cdot 10^{-5}$ at a distance of 1\,kpc.
 Since the computational cost of such searches scales with $f^2$, low-frequency searches are actually much more efficient and would allow very deep searches of the OGO data, further increasing the competitiveness.
 Note, however, that for low-frequency pulsars the ellipticities required to achieve detectable GW signals can be very high, possibly mostly in the unphysical regime.
 On the other hand, for ``transient CW''-type signals~\cite{Prix2011}, low-frequency pulsars might be the strongest emitters, even with realistic ellipticities.

 \subsection{Other sources}

 Many (indirect) observational evidences exist for stellar mass BHs, which are the end stages of heavy star evolution, as well as for supermassive BHs, the result of accretion and galactic mergers throughout the cosmic evolution, in galactic nuclei.
 On the other hand, there is no convincing evidence so far for a BH of an intermediate mass in the range of $10^2-10^4\msun$.
 These IMBHs might, however, still exist in dense stellar clusters~\cite{Miller2004, Pasquato2010}.
 Moreover, stellar clusters could be formed as large, gravitationally bound groups, and collision of two clusters would produce inspiralling binaries of IMBHs~\cite{IMBH_pau2006,IMBH_pau2009}.

 The ISCO frequency of the second orbital harmonic for a $300 \msun$-$300 \msun$ system is about 7\,Hz, which is outside the sensitivity range of aLIGO.
 Still, those sources could show up through the higher harmonics (the systems are expected to have non-negligible eccentricity) and through the merger and ring-down gravitational radiation~\cite{Fregeau2006,Mandel2008,Yagi2012}.
 The ground-based LIGO and VIRGO detectors have already carried out a first search for IMBH signals in the $100\msun$ to $450\msun$ mass range~\cite{Abadie2012_imbh}.

 With its better low-frequency sensitivity, OGO can be expected to detect a GW signal from the inspiral of a $300 \msun$-$300 \msun$ system in a quasicircular orbit up to a distance of approximately 245\,Mpc, again using Eq.~(\ref{eq:Dhorizon}). 
 This gives the potential for discovery of such systems and for estimating their physical parameters.

 As for other advanced detectors, unmodeled searches (as opposed to the matched-filter CBC and CW searches; see Ref.~\cite{Abadie2012_burst} for a LIGO example) of OGO data have the potential for detecting many other types of gravitational wave sources, including, but not limited to, supernovae and cosmic string cusps.
 However, as in the case for IMBHs, the quantitative predictions are hard to produce due to uncertainties in the  models.

\section{Summary and Outlook}
 \label{S:Summary}
 In this paper, we have presented for the first time a three-dimensional gravitational wave detector in space, called the Octahedral Gravitational wave Observatory (OGO).
 The detector concept employs displacement-noise free interferometry (DFI), which is able to cancel some of the dominant noise sources of conventional GW detectors.
 Adopting the octahedron shape introduced in Ref.~\cite{chen2006}, we put spacecraft in each corner of the octahedron.
 We considered a LISA-like receiver-transponder configuration and found multiple combinations of measurement channels, which allow us to cancel both laser frequency and acceleration noise.
 This new three-dimensional result generalizes the Mach-Zehnder interferometer considered in Ref.~\cite{chen2006}.

 We have identified a possible halolike orbit around the Lagrange point L1 in the Sun-Earth system that would allow the octahedron constellation to be stable enough.
 However, this orbit limits the detector to an arm length of $\approx1400$ km.

 Much better sensitivity and a richer astrophysical potential are expected for longer arm lengths.
 Therefore, we also looked for alternative orbits and found a possible alternative allowing for $\approx 2\cdot10^{9}$ m arms, but is is not clear yet if this would be stable enough.
 Future studies are required to relax the equal-arm-length assumption of our DFI solutions, or to determine a stable, long-arm-length constellation.

 Next, we have computed the sensitivity of OGO-like detectors -- and have shown that the noise-cancelling combinations also cancel a large fraction of the GW signal at low frequencies.
 The sensitivity curve therefore has a characteristic slope of $f^{-2}$ at the low-frequency end.

 However, the beauty of this detector is that it is limited by a single noise source at all frequencies: shot noise.
 Thus, any reduction of shot noise alone would improve the overall sensitivity.
 This could, in principle, be achieved with DECIGO-like cavities, squeezing or other advanced technologies.
 Also, OGO does not require drag-free technology and has moderate requirements on other components so that it could be realized with technology already developed for LISA Pathfinder and eLISA.

 When comparing a DFI-enabled OGO with a detector of similar design, but with standard TDI, we find that at $\approx1400$ km, the same sensitivity could be reached by a TDI detector with very modest acceleration noise requirements.

 However, at longer arm lengths DFI becomes more advantageous, reaching the same sensitivity as TDI under LISA requirements but without drag-free technology and clock transfer, at $\approx 2\cdot10^{9}$ m.
 Such a DFI detector would have its best frequency range between LISA and DECIGO, with peak sensitivity better than LISA and approaching DECIGO without the latter mission concept's tight acceleration noise requirements and with no need for cavities.

 Finally, we have assessed the scientific potentials of OGO, concentrating on the less promising, but more mature short-arm-length version.
 We estimated the event rates for coalescing binaries, finding that OGO is better than initial LIGO, but does not reach the level of advanced LIGO.
 Any binary detected with both OGO and aLIGO could be localized in the sky with very high accuracy.

 Also, the three-dimensional satellite constellation and number of independent links makes OGO an interesting mission for detection of the stochastic GW background or hypothetical additional GW polarizations.
 Further astrophysically interesting sources such as low-frequency pulsars and IMBH binaries also lie within the sensitive band of OGO, but again the sensitivity does not reach that of aLIGO.

 However, we point out that the improvement in the low-frequency sensitivity with increasing arm length happens faster for DFI as compared to the standard TDI.
 Therefore, searching for stable three-dimensional (octahedron) long-baseline orbits could lead to an astrophysically much more interesting mission.

 Regarding possible improvements of the presented setup, there are several possibilities to extend and improve the first-order DFI scheme presented here.
 One more spacecraft could be added in the middle, increasing the number of usable links.
 Breaking the symmetry of the octahedron could modify the steep response function at low frequencies.
 This should be an interesting topic for future investigations.

 In principle, the low-frequency behavior of OGO-like detectors could also be improved by more advanced DFI techniques such as introducing artificial time delays~\cite{Somiya2007a, Somiya2007b}.
 This would result in a three-part power law less steep than the shape derived in Sec.~\ref{sec:transfer_function}.
 On the other hand, this would also introduce a new source of time delay noise.
 Therefore, such a modification requires careful investigation.

\section{Acknowledgments}
 We would like to thank Gerhard Heinzel for very fruitful discussions, Albrecht R\"{u}diger for carefully reading through the paper and helpful comments, Sergey Tarabrin for discussions on the optical layout, Masaki Ando for kindly sharing DECIGO simulation tools and Guido M\"uller for helpful comments on the final manuscript.
 Moreover, Berit Behnke, Benjamin Knispel, Badri Krishnan, Reinhard Prix, Pablo Rosado, Francesco Salemi, Miroslav Shaltev and others helped us with their knowledge regarding the astrophysical sources.
 We would also like to thank the anonymous referee for very insightful and detailed comments on the original manuscript.
 The work of the participating students was supported by the International Max-Planck Research School for Gravitational Waves (IMPRS-GW) grant.
 The work of S. B. and Y. W. was partially supported by DFG Grant No. SFB/TR 7 Gravitational Wave Astronomy and DLR (Deutsches Zentrum f\"{u}r Luft- und Raumfahrt).
 Furthermore, we want to thank  the Deutsche Forschungsgemeinschaft (DFG) for funding the Cluster of Excellence QUEST -- Centre for Quantum Engineering and Spacetime Research.
 We thank the LIGO Scientific Collaboration (LSC) for supplying the LIGO and aLIGO sensitivity curves.
 Finally, we would like to emphasize that the idea of a three-dimensional GW detector in space is the result of a student project from an IMPRS-GW lecture week.
 This document has been assigned LIGO document number and LIGO-P1300074 and AEI-preprint number AEI-2013-261.

\appendix

\section{Details on calculating the displacement and laser noise free combinations}
 \label{S:Appendix}

 Here we will give details on building the displacement (acceleration) and laser noise free configurations.
 The derivations closely follow the method outlined in~\cite{TintoDhurandhar}.
We want to find the generators solving Eq.~(\ref{E:TDIAcc_FinalEq}), so called reduced generators because they correspond to the reduced set
$( q_{BC}, q_{CE}, q_{DB}, q_{DC}, q_{DF}, q_{EF} )$.  We start with  building  the ideal $Z$:
 \begin{align}
  Z = \left\{
  \begin{array}{rcl}
  f_1 &=& (\mathcal{D}-1)^2 \\
  f_2 &=& (\mathcal{D}-1)\mathcal{D} \\
  f_3 &=& (1-\mathcal{D})(\mathcal{D}-1) \\
  f_4 &=& (\mathcal{D}-1)((1-\mathcal{D})\mathcal{D}-1) \\
  f_5 &=& \mathcal{D}-1  \\
  f_6 &=& \mathcal{D}-1  \\
  \end{array}
  \right.\,.
  \label{E:TDIAcc_ideal}
 \end{align}
 The corresponding Gr\"obner basis to this ideal is:
 \begin{equation}
  \mathcal{G}=\{ g_1 = \mathcal{D} - 1 \}.
  \label{E:E:TDIAcc_Groebner}
 \end{equation}

 The connection between $f_i$ and $g_j$ is defined by two transformation matrices
 \begin{eqnarray}
  d & =& \left(
  \begin{array}{c}
  \mathcal{D}-1 \\
  \mathcal{D} \\
  1-\mathcal{D} \\
  (1-\mathcal{D})\mathcal{D}-1 \\
  1 \\
  1 \\
  \end{array}
  \right)
  \end{eqnarray}
  and $c$ with (at least) two possible solutions
  \begin{equation}
  c^{(1)} = \left( 0 \ 0 \ 0 \ 0 \ 1 \ 0 \right)\; \textrm{or} \; c^{(2)} = \left( 0 \ 0 \ 0 \ 0 \ 0 \ 1 \right).
  \end{equation}
 The resulting basis is not unique and not necessarily independent.
 The first 6 reduced generators are given by the row vectors of the matrix $A^{(1)} = a^{(1)}_i = I - d\cdot c^{(1)}$ :
 \begin{subequations}
 \begin{align}
  a^{(1)}_1 & = \left\{ 1 , 0 , 0 , 0 , 0 , 1-\mathcal{D} \right\}, \\
  a^{(1)}_2 & = \left\{ 0 , 1 , 0 , 0 , 0 , -\mathcal{D} \right\}, \\
  a^{(1)}_3 & = \left\{ 0 , 0 , 1 , 0 , 0 , (\mathcal{D}-1)\mathcal{D} \right\}, \\
  a^{(1)}_4 & = \left\{ 0 , 0 , 0 , 1 , 0 , 1+(\mathcal{D}-1)\mathcal{D} \right\}, \\
  a^{(1)}_5 & = \left\{ 0 , 0 , 0 , 0 , 1 , -1 \right\}, \\
  a^{(1)}_6 & = \left\{ 0 , 0 , 0 , 0 , 0 , 0 \right\}.
 \end{align}
 \end{subequations}
 These reduced generators correspond directly to values for $( q_{BC}, q_{CE}, q_{DB}, q_{DC}, q_{DF}, q_{EF} )$.
 As the Gr\"obner basis contains only one element, we cannot form other generator from $S$-polynomial.

 We can form 6 other generators using $c^{(2)}$ instead of $c^{(1)}$.
 After applying those generators we have the following acceleration-free combinations:
 \begin{subequations}
 \begin{align}
  a^{(1)}_1 s^{n} & = 2 (p_{B} - p_{C} + p_{E} - p_{F} + \mathcal{D}(-p_{A} + p_{B} - p_{D} + p_{E} \nonumber \\ &
+ (p_{B} - p_{C} + p_{E} - p_{F}) q_{BA})), \\
  a^{(1)}_2 s^{n} & = -2 \mathcal{D} (p_{A} + p_{D} + p_{C} (-1 + q_{BA}) + p_{F} (-1 + q_{BA}) \nonumber \\ &
- (p_{B} + p_{E}) q_{BA}), \\
  a^{(1)}_3 s^{n} & = 2 \mathcal{D} ((1 + \mathcal{D}) p_{A} + p_{D} - p_{E} - \mathcal{D} (p_{C} - p_{D} + p_{F}) \nonumber \\ &
 + p_{B} (-1 + q_{BA}) - (p_{C} - p_{E} + p_{F}) q_{BA}), \\
  a^{(1)}_4 s^{n} & = 2 (p_{B} - p_{C} + p_{E} + \mathcal{D}^2 (p_{A} - p_{C} + p_{D} - p_{F}) \nonumber \\ &
 - p_{F} + \mathcal{D} (p_{B} - p_{C} + p_{E} - p_{F}) q_{BA}), \\
  a^{(1)}_5 s^{n} & = 2 \mathcal{D} (p_{A} + p_{D} + p_{B} (-1 + q_{BA}) + p_{E} (-1 + q_{BA}) \nonumber \\ &
 - (p_{C} + p_{F}) q_{BA}), \\
  a^{(1)}_6 s^{n} & = 2 \mathcal{D} (p_{B} - p_{C} + p_{E} - p_{F}) q_{BA},
 \end{align}
 \end{subequations}
where $s^{n}_{IJ}$ are given by Eq.~(\ref{E:MesN}). Note that we have a free (polynomial) function of delay $q_{BA}$ which we can choose arbitrary.
We will omit subscripts $BA$ and use $q\equiv q_{BA}$. The arbitrariness of this function implies that terms which contain $q$ and terms free of $q$
are two independent sets of generators. We will keep $q$ until we obtain laser noise free combinations, and then split each generator in two.
 After some analysis only two out of six acceleration free generators are independent, so we can rewrite them as
 \begin{subequations}
 \begin{align}
  s_1 &= y_{12} + \mathcal{D}(y_{13} + qy_{12}),\\
  s_3 &= -y_{13} + \mathcal{D}(y_{12} -y_{13}) + qy_{12},\\
  s_4 &= y_{12} + \mathcal{D}q y_{12} + \mathcal{D}^2(y_{12} - y_{13}),\\
  s_2+s_5 &= y_{12} - 2y_{13},\label{E:s2ps5}\\
  s_2-s_5 &= (2q-1)y_{12},\label{E:s2ms5}\\
  s_6 &= q y_{12},
 \end{align}
 \end{subequations}
 where
 \begin{eqnarray}
 s_1 &=& \frac{a^{(1)}_1 s^{n}}{2}, s_2 = -\frac{\mathcal{D}^{-1}( a^{(1)}_2 s^{n})}{2}, s_3 = \frac{\mathcal{D}^{-1}( a^{(1)}_3 s^{n})}{2} \nonumber \\
 s_4 &=& \frac{a^{(1)}_4 s^{n}}{2}, s_5 = \frac{\mathcal{D}^{-1}( a^{(1)}_5 s^{n})}{2} , s_6 = \frac{\mathcal{D}^{-1}( a^{(1)}_6 s^{n})}{2}
\end{eqnarray}
and
\begin{equation}
y_{12} = p_B+p_E-p_C-p_F,\, y_{13} = p_B+p_E-p_A-p_D\,.
\end{equation}
We have introduced the inverse delay operator,
$ \mathcal{D}^{-1}$,
 for mathematical convenience, which obeys $\mathcal{D}\mathcal{D}^{-1} = \mathds{1}$. One can easily get rid of it by applying the delay operator on both sides.
 The final result will not contain the operator $\mathcal{D}^{-1}$.
 Next we use Eqs.~(\ref{E:s2ps5}) and (\ref{E:s2ms5}) to express $y_{12}, y_{13}$ and eliminate them from the other equations.
 The resulting combinations that eliminate both acceleration and laser noise are
 \begin{subequations}
 \begin{align}
  &(1-2q)s_1 + (-1-2\mathcal{D}q)s_2  + (1+\mathcal{D})s_5\\
  &(1-2q)s_3 + \mathcal{D}(q-1)s_2 + (-1+2q+q\mathcal{D})s_5\\
  &(1-2q)s_4 - (1+ \mathcal{D}q)(s_2-s_5) - \mathcal{D}^2((1-q)s_2 - qs_5)\\
  &(1-2q)s_6 - q(s_2 - s_5).
 \end{align}
 \end{subequations}
Out of these solutions we obtain seven independent generators which we have rewritten in the final form similar to the $Y$-equations from Sec.~\ref{S:TDI}.
They are explicitly given by Eqs.~(\ref{E:S1})--(\ref{E:S7}).

\end{document}